\documentclass[aps,pre,twocolumn,showpacs]{revtex4}
\usepackage{amsmath}
\usepackage{amsfonts}
\usepackage{bm}
\usepackage[final]{epsfig}

\newcommand{\angint}{\int_0^{2\pi}\negthickspace\negthickspace}

\begin{document}
\title{Eigenfunction approach to the persistent random walk in two dimensions}
\author{Christian Bracher}
\email{cbracher@fizz.phys.dal.ca}
\affiliation{Department of Physics and Atmospheric Science, Dalhousie University, Halifax, N.S.\ B3H~3J5, Canada}
\date{\today}

\begin{abstract}
The Fourier--Bessel expansion of a function on a circular disc yields a simple series representation for the end-to-end probability distribution function $w(R,\phi)$ encountered in a planar persistent random walk, where the direction taken in a step depends on the relative orientation towards the preceding step.  For all but the shortest walks, the proposed method provides a rapidly converging, numerically stable algorithm that is particularly useful for the precise study of intermediate-size chains that have not yet approached the diffusion limit.   As a practical application, we examine the force-extension diagram of various planar polymer chains.  With increasing joint stiffness, a marked transition from rubber-like behaviour to a form of elastic response resembling that of a flexible rod is observed.
\end{abstract}

\maketitle

\section{Introduction}
\label{sec:Intro}

Today, the concept of the random walk is an ubiquitous tool in the physical sciences, and a multitude of different mathematical and physical problems have been subsumed under this catch-all phrase \cite{Shlesinger1984a,Weiss1994a}.  Originally, the term has been coined by Pearson \cite{Pearson1905a} who was interested in the probability distribution $w_P(R)$ for the end-to-end distance in a chain of $N$ randomly chosen steps of equal length $l$ in a plane.  This article concerns a random walk process that is closely related to the initial formulation of the problem:  While Pearson's general setup is maintained, we introduce an angular bias into the relative orientation of successive steps that is governed by a probability distribution $p(\theta)$.  Known as the continuous persistent random walk in the plane, this modification continues to attract considerable theoretical interest \cite{Weiss1994a,Masoliver1993a,Tojo1996a,Larralde1997a,Wu2000a}.  Practical applications of the planar walk cover a wide range including light scattering, where sums of random complex numbers are studied \cite{Barakat1974a}, its related use in crystal structure analysis \cite{Shmueli1990a,Weiss1994a,Shmueli1995a}, and surface diffusion processes.  The model of the persistent walk examined here is of particular interest to the configurational statistics of polymers \cite{Flory1969a}, where the inherent stiffness of chemical bonds restricts the relative orientation of adjacent monomers, and obviously for two-dimensional locomotion problems in biology and medicine \cite{Nossal1974a,Nossal1974b,Nossal1983a,Wu2000a}.

In several of these publications \cite{Tojo1996a,Larralde1997a,Wu2000a,Flory1969a}, only moments of the probability distribution function $w(R,\phi)$ were calculated, while it appeared desirable to the authors to have access to $w(R,\phi)$ beyond the diffusion limit $N\rightarrow\infty$, where this function ultimately takes on a simple Gaussian shape.  Unfortunately, the search for explicit solutions even for Pearson's original problem has remained largely elusive.  While some recent contributors resorted to numerical Monte Carlo simulation \cite{Tojo1996a}, in a pioneering work Barakat \cite{Barakat1973a} realized that expansion  into a discrete Fourier series offers a superior alternative in computing the radial distribution function $w(R)$ of isotropic random walks.  The corresponding Fourier--Bessel series solution for the Pearson random  walk is equally due to Barakat \cite{Barakat1974a}, and subsequently has been advantageously applied in crystallographical analysis \cite{Shmueli1990a,Weiss1994a}.  In this article, we extend the Fourier analysis to the persistent random walk problem characterized by an anisotropic end-to-end distribution function $w(R,\phi)$ and find excellent numerical convergence for reasonably smooth angular distribution functions $p(\theta)$, except for the shortest walks.  This renders the method particularly suited for the study of intermediate-size chains whose behaviour conspicuously deviates from the diffusion limit.  Our derivation of the series representation rests on rather elementary methods.  (A more complicated general analysis based on the characteristic function $c(\omega,\chi)$ has been put forward by Weiss and Shmueli \cite{Weiss1987a}.  When applied to the persistent random walk, it implicitly yields the same result.)

Let us briefly outline the organization of this article:  In Section~\ref{sec:Theory}, we develop the theory of the persistent planar walk.  Starting with moments of the distribution, we proceed to derive the characteristic function $c(\omega,\chi)$ in form of a matrix product, which we use to deduce an integral representation of $w(R,\phi)$ reminiscent of Kluyver's solution to Pearson's problem \cite{Kluyver1906a}.  Finally, we establish the Fourier--Bessel eigenfunction expansion of this distribution.  Numerical results obtained with our method are presented in Section~\ref{sec:Example}, where we study the convergence behaviour of the series and display the frequently distinctive shapes of the distribution function obtained for several choices of the angular weight $p(\theta)$.  Some possible extensions of the theory are discussed in Section~\ref{sec:Diss}.  Finally, a proof of the ``projection theorem'' that forms the basis of our proposed series expansion is sketched in Appendix~\ref{sec:Projection}, while Appendix~\ref{sec:Moments} covers the calculation of moments in the Fourier--Bessel scheme.

\section{Theory of the~2D~persistent walk}
\label{sec:Theory}

\begin{figure}
\centerline{\includegraphics[draft=false,width=0.9\columnwidth]{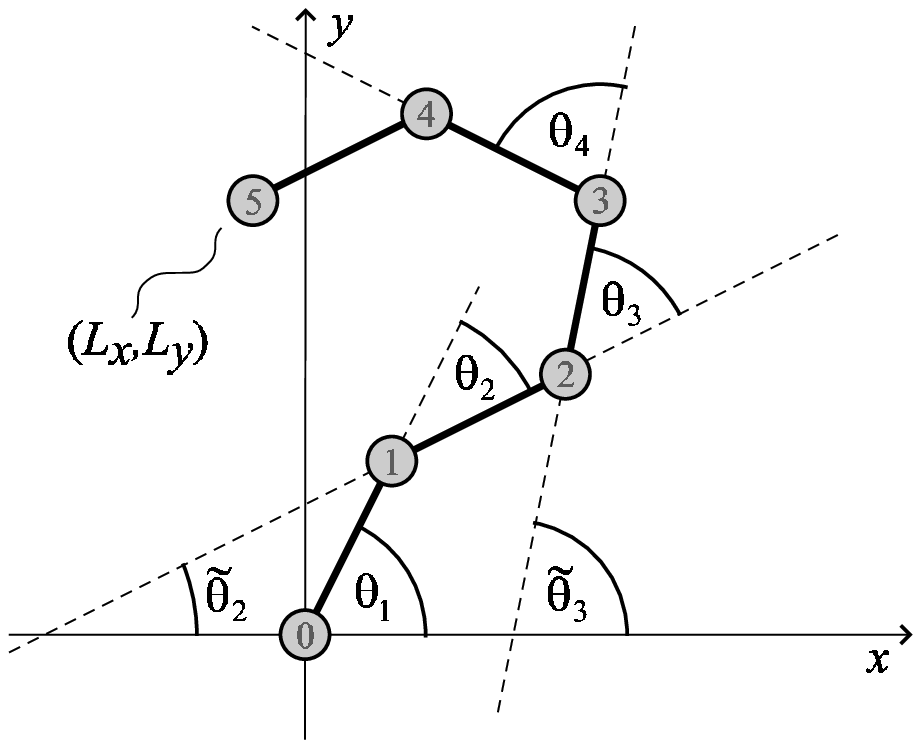}}
\caption{The geometry of the planar random walk.  The planar walk is described by the set of relative angles $\{ \theta_1,\theta_2,\ldots,\theta_N \}$ formed by adjacent steps, or alternatively by the set of absolute angles $\{ \tilde\theta_1,\tilde\theta_2,\ldots,\tilde\theta_N \}$ of the segments with respect to the $x$--axis.  (Note that $\theta_1 = \tilde\theta_1$.)
\label{fig:Angles}}
\end{figure}

We start out with some basic conventions.  For simplicity of notation, in this paper we deal exclusively with a planar persistent random walk built of $N$ steps with uniform length $l$, but the generalization of our theory to incorporate individually different step lengths is straightforward.  The walk is then fully characterized by the set of relative angles $\{ \theta_1,\theta_2,\ldots,\theta_N \}$ at the links of successive segments, where the initial angle $\theta_1$ is measured relative to the $x$--axis of our coordinate system (Figure~\ref{fig:Angles}).  In practice, the absolute angles $\tilde\theta_\nu$ formed between the $\nu$.th step and the $x$--axis as direction of reference are equally important, and they form an alternative description of the chain.  As easily inferred from Figure~\ref{fig:Angles}, both sets of angles are interrelated via:
\begin{equation}
\label{eq:Theory0.1}
\tilde\theta_\nu -\tilde\theta_{\nu-1} = \theta_\nu \quad\longleftrightarrow\quad \tilde\theta_\nu = \sum_{k=1}^\nu \theta_k \;.
\end{equation}
Since we are mainly interested in the end-to-end distribution function $w(R,\phi)$, we start the random walk from the origin of our coordinate system.  We denote its endpoint by $(L_x,L_y)$ in cartesian and $(R,\phi)$ in polar coordinates, respectively.

For the persistent walk, the probability of a certain configuration $p(\theta_1,\theta_2,\ldots,\theta_N)$ depends only on the relative angles $\theta_\nu$ formed by adjacent steps.  Assuming that these angles are not correlated, and that all links are equivalent (including the initial ``joint'' between the first step and the $x$--axis), this probability reduces to a product of independent factors: $p(\theta_1,\ldots,\theta_N) = p(\theta_1) \cdot\ldots\cdot p(\theta_N)$, where the angular probability distribution $p(\theta) = p(\theta + 2\pi)$ is a periodic function of the change of direction $\theta$.  This suggests that a Fourier expansion of the function $p(\theta)$ will be convenient \cite{Nossal1974a,Nossal1974b}:
\begin{equation}
\label{eq:Theory0.2}
p(\theta) = \frac 1{2\pi} \sum_\nu {\rm e}^{-{\rm i}\nu\theta} p_\nu \quad\longleftrightarrow\quad
p_\nu = \angint {\rm d}\theta \, {\rm e}^{{\rm i}\nu\theta} p(\theta) \;.
\end{equation}
Proper normalization of $p(\theta)$ requires $p_0 = 1$; the norm of all other (generally complex) angular Fourier components is then less than unity, $|p_\nu| < 1$ for $\nu \neq 0$.  (Equality may occur only for rigidly linked chains with $p(\theta) = \delta(\theta -\theta_0)$ which we exclude here.)  With these definitions, we are now set to determine quantities of interest for the walk.

\subsection{Persistence length and diffusion constant}
\label{sec:Theory1}

Our first investigation concerns the average position $(\langle L_x \rangle_N, \langle L_y \rangle_N)$ of the chain end and its mean square displacement $\langle L_x^2 + L_y^2 \rangle_N$ as a function of the number of steps $N$.  Both quantities are accessible in a simple calculation that introduces some techniques that will be commonplace below.  It is convenient to interpret vectors in the $x-y$--plane as complex numbers; with this convention, the vector formed by the $\nu$.th element $(l_x^{(\nu)}, l_y^{(\nu)})$ is characterized by its absolute angle $\tilde\theta_\nu$ with respect to the $x$--axis (Figure~\ref{fig:Angles}): $l_x^{(\nu)} + {\rm i}l_y^{(\nu)} = l\,{\rm e}^{{\rm i}\tilde\theta_\nu}$.  The position of the chain end is simply the sum of all constituent step vectors, $L_x + {\rm i}L_y = l \sum_{\nu =1}^N {\rm e}^{{\rm i}\tilde\theta_\nu}$, and its properly weighed average over all configurations of a $N$--element walk correspondingly reads:
\begin{align}
\nonumber
\left\langle L_x + {\rm i}L_y \right\rangle_N &= l \sum_{\nu=1}^N \angint {\rm d}\theta_1  p(\theta_1) \cdots \angint {\rm d}\theta_N p(\theta_N) \,{\rm e}^{{\rm i}\tilde\theta_\nu} \\
&= l \sum_{\nu = 1}^N p_0^{N-\nu} p_1^\nu  = l \frac{p_1 \left( 1 - p_1^N \right)}{1 - p_1} \;.
\label{eq:Theory1.1}
\end{align}
Here, we successively used the resolution of the absolute angle $\tilde\theta_\nu$ into relative angles $\theta_k$ (\ref{eq:Theory0.1}), the definition of the angular Fourier components in (\ref{eq:Theory0.2}), and finally the normalization condition $p_0 = 1$.  We infer that the average displacement of the chain end depends only on the first Fourier component of $p(\theta)$, and if $p_1$ vanishes, the distribution is centered around the origin for all $N$.  Unless the chain is absolutely rigid ($|p_1| = 1$), the mean end-to-end vector approaches a well-defined limit as $N\rightarrow \infty$:
\begin{equation}
\label{eq:Theory1.2}
\lim_{N\rightarrow \infty} \left\langle L_x + {\rm i}L_y \right\rangle_N = l\, \frac{p_1}{1-p_1} \;,
\end{equation}
which generalizes the notion of the persistence length of the chain put forward in Ref.~\cite{Flory1969a}.  For comparison, we also state this result using real parameters.  Introducing $p_1 = \alpha {\rm e}^{{\rm i}\beta}$, where $\beta$ represents the average ``slant'' angle in the angular distribution $p(\theta)$ \cite{Tojo1996a,Larralde1997a}, one finds in the long-chain limit:
\begin{align}
\label{eq:Theory1.3a}
\left\langle L_x \right\rangle_\infty &= \frac{l \alpha (\cos\beta - \alpha)}{1-2\alpha\cos\beta+\alpha^2} \;,\\
\label{eq:Theory1.3b}
\left\langle L_y \right\rangle_\infty &= \frac{l \alpha \sin\beta}{1-2\alpha\cos\beta+\alpha^2} \;,
\end{align}
which yields for the average absolute displacement $\langle R \rangle = (\langle L_x \rangle^2 + \langle L_y \rangle^2)^{1/2}$ as $N\rightarrow \infty$:
\begin{equation}
\label{eq:Theory1.4}
\langle R \rangle_\infty = \frac{l\alpha}{\sqrt{1-2\alpha\cos\beta+\alpha^2}} \;.
\end{equation}
This simple expression might be useful in the interpretation of experimental data \cite{Wu2000a}.

The determination of the mean square displacement $\langle R^2 \rangle_N = \langle L_x^2 + L_y^2 \rangle_N$ is only slightly more involved.  In the complex notation, $R^2 = (L_x + {\rm i}L_y)(L_x - {\rm i}L_y) = l^2 \sum_{\nu,\mu =1}^N {\rm e}^{{\rm i}(\tilde\theta_\nu - \tilde\theta_\mu)}$, and the same series of arguments we used in (\ref{eq:Theory1.1}) yields:
\begin{align}
\nonumber
\left\langle R^2 \right\rangle_N &= l^2 \sum_{\nu,\mu=1}^N \angint {\rm d}\theta_1  p(\theta_1) \cdots \angint {\rm d}\theta_N p(\theta_N) \,{\rm e}^{{\rm i}(\tilde\theta_\nu-\tilde\theta_\mu)} \\
&= Nl^2 + l^2 \sum_{\delta = 1}^N (N-\delta) \left( p_1^\delta + p_{-1}^\delta \right) \;,
\label{eq:Theory1.5}
\end{align}
where we exploited $\tilde\theta_\nu-\tilde\theta_\mu = \theta_{\mu +1}+\ldots+\theta_\nu$ $(\nu>\mu)$ and sorted the terms in the double sum according to their ``distance'' $\delta =|\nu-\mu|$.  The arithmetic-geometric progression in (\ref{eq:Theory1.5}) is easily evaluated \cite{Gradshteyn1980a}:
\begin{equation}
\label{eq:Theory1.6}
\left\langle R^2 \right\rangle_N = \frac{Nl^2}2\,\frac{1+p_1}{1-p_1} - l^2\,\frac{p_1(1-p_1^N)}{(1-p_1)^2} + \text{c.c.}
\end{equation}
(A very similar result is obtained for the mean square displacement in the three-dimensional freely rotating chain \cite{Flory1969a}.)  Since the complex conjugate term is formally obtained by replacing $p_1$ with $p_{-1}$ in (\ref{eq:Theory1.6}), the mean square displacement, like the absolute displacement (\ref{eq:Theory1.1}), depends only on the first angular Fourier component $p_{\pm 1}$.  For $p_1=0$, (\ref{eq:Theory1.6}) reduces to $\left\langle R^2 \right\rangle_N = Nl^2$, the trivial result obtained in Pearson's walk.  In the limit $N\rightarrow\infty$, we arrive at the following diffusion constant (characteristic ratio) $D_N = \langle R^2 \rangle_N / N$ for the persistent walk:
\begin{equation}
\label{eq:Theory1.7}
D_\infty = \frac{l^2 (1 - p_1p_{-1})}{(1-p_1)(1-p_{-1})} = \frac{l^2(1-\alpha^2)}{1-2\alpha\cos\beta+\alpha^2} \;.
\end{equation}
The asymmetry parameters $\alpha$ and $\beta$ of the angular distribution $p(\theta)$ thus determine the persistence length $\langle R \rangle_\infty$ (\ref{eq:Theory1.4}) and the diffusion constant $D_\infty$ of the walk and vice versa.  We point out that for $\beta\neq 0$, with increasing $\alpha$ $D_\infty$ starts to diminish from a certain stiffness of the joint and actually vanishes as $\alpha\rightarrow1$.  This property reflects the fact that the walk then tends to move around a circle, and no net transport takes place.  (An alternative derivation of (\ref{eq:Theory1.7}) has been given by Larralde \cite{Larralde1997a}.)  We conclude this section by noting that higher momenta of the distribution function $w(R,\phi)$ evidently are available from similar calculations.  However, in practice they are more easily evaluated from a Fourier--Bessel series representation (Appendix~\ref{sec:Moments}).

\subsection{The characteristic function}
\label{sec:Theory2}

In the previous section, we were able to determine the moments $\langle L_x + {\rm i}L_y\rangle_N$ (\ref{eq:Theory1.1}) and $\langle R^2 \rangle_N$ (\ref{eq:Theory1.5}) of the probability distribution $w(R,\phi)$ by summation of some geometric series.  We now condense the ideas implicitly introduced there into a single, coherent formalism and employ it towards the corresponding characteristic function $c(\omega,\chi) = \langle \exp({\rm i}\boldsymbol\omega\cdot\mathbf L) \rangle_N$ of the persistent walk.

So far, we evaluated averages of the form $\langle \exp({\rm i}\tilde\theta_\nu) \rangle$ and $\langle \exp[{\rm i}(\tilde\theta_\nu-\tilde\theta_\mu)] \rangle$, respectively.  Both expressions may be viewed as special cases of the more general average $\langle \exp({\rm i}\sum_\nu \lambda_\nu \tilde\theta_\nu) \rangle$, where $\{ \lambda_1,\ldots,\lambda_N\}$ is a set of $N$ integers and the sum covers the range $1\leq\nu \leq N$.  In view of the relation $\tilde\theta_\nu = \theta_1 + \ldots + \theta_\nu$ (\ref{eq:Theory0.1}) between absolute and relative angles, its expression in terms of the angular Fourier coefficients $p_\lambda$ (\ref{eq:Theory0.2}) is straightforward:
\begin{equation}
\label{eq:Theory2.1}
\left\langle \exp\left( {\rm i}\sum\nolimits_\nu \lambda_\nu \tilde\theta_\nu \right) \right\rangle =
p_{\lambda_N}p_{\lambda_N + \lambda_{N-1}}\cdot\ldots\cdot p_{\lambda_N + \ldots +\lambda_2 + \lambda_1} \;.
\end{equation}
The significance of this development is as follows.  Consider a functional $Q(\tilde\theta_1,\ldots,\tilde\theta_N)$ of the random walk that depends only on the absolute angles $\tilde\theta_\nu$ of the segments.  In order to be meaningful, such a function must be periodic in each of those angles, $Q(\ldots,\tilde\theta_\mu + 2\pi,\ldots) = Q(\ldots,\tilde\theta_\mu,\ldots)$, which in turn implies that it may be represented by a discrete Fourier series:
\begin{equation}
\label{eq:Theory2.2}
Q(\tilde\theta_1,\ldots,\tilde\theta_N) = \sum_{\mu_1\cdots\mu_N} Q_{\mu_1,\ldots,\mu_N}
{\rm e}^{{\rm i}(\mu_1\tilde\theta_1 + \ldots + \mu_N\tilde\theta_N)} \;,
\end{equation}
where integration yields the Fourier coefficient $Q_{\mu_1,\ldots,\mu_N}$:
\begin{multline}
\label{eq:Theory2.3}
Q_{\mu_1,\ldots,\mu_N} = \frac1{(2\pi)^N} \angint {\rm d}\tilde\theta_1 \cdots \angint {\rm d}\tilde\theta_N
\\ \times\, {\rm e}^{-{\rm i}(\mu_1\tilde\theta_1 + \ldots + \mu_N\tilde\theta_N)}  p(\tilde\theta_1,\ldots,\tilde\theta_N) \;.
\end{multline}
(It proves convenient here to employ a Fourier expansion scheme that differs from the angular decomposition (\ref{eq:Theory0.2}) in its sign convention and normalization.)  By combination of (\ref{eq:Theory2.1}) and (\ref{eq:Theory2.3}), we find the average value for the functional $\langle Q(\tilde\theta_1,\ldots,\tilde\theta_N) \rangle$ in terms of Fourier coefficients:
\begin{align}
\nonumber
\left\langle Q(\tilde\theta_1,\ldots,\tilde\theta_N) \right\rangle &=
\sum_{\mu_1\cdots\mu_N} Q_{\mu_1,\ldots,\mu_N} p_{\mu_N}\cdot\ldots\cdot p_{\mu_N + \ldots + \mu_1}
\\
&= \begin{aligned}[t]
     \sum_{\nu_1\cdots\nu_N} & Q_{\nu_1-\nu_2,\nu_2-\nu_3,\ldots,\nu_{N-1}-\nu_N,\nu_N}  \\
     & \times \, p_{\nu_N}p_{\nu_{N-1}}\cdot\ldots\cdot p_{\nu1} \;.
  \end{aligned}
\label{eq:Theory2.4}
\end{align}
In the second line of this equation, we employed an alternative indexing scheme $\{ \nu_1,\ldots,\nu_N \}$ defined via:
\begin{equation}
\label{eq:Theory2.5}
\nu_k = \mu_k + \ldots + \mu_N \quad\longleftrightarrow\quad \mu_k = \nu_k - \nu_{k+1} \;,
\end{equation}
$(\mu_N = \nu_N)$.  Its advantages will become obvious below.

We are particularly interested in the special case where the functional $Q(\tilde\theta_1,\ldots,\tilde\theta_N)$ reduces to a product of $N$ equivalent factors $q(\tilde\theta_\nu)$:  $Q(\tilde\theta_1,\ldots,\tilde\theta_N) = \prod_\nu q(\tilde\theta_\nu)$.  Under these circumstances, the Fourier components $Q_{\mu_1,\ldots,\mu_N}$ will factorize as well, and we have $Q_{\mu_1,\ldots,\mu_N} = q_{\mu_1}\cdot\ldots\cdot q_{\mu_N}$, where:
\begin{equation}
\label{eq:Theory2.6}
q(\tilde\theta) = \sum_\mu {\rm e}^{{\rm i}\mu\tilde\theta} q_\mu \quad\longleftrightarrow\quad
q_\mu = \frac1{2\pi} \angint {\rm d}\tilde\theta \,{\rm e}^{-{\rm i}\mu\tilde\theta} q(\tilde\theta) \;.
\end{equation}
We now exploit these properties in order to simplify the average (\ref{eq:Theory2.4}).  Upon inserting (\ref{eq:Theory2.6}), we rearrange the summations involved:
\begin{multline}
\label{eq:Theory2.7}
\left\langle Q(\tilde\theta_1,\ldots,\tilde\theta_N) \right\rangle = \sum_{\nu_N} q_{\nu_N}p_{\nu_N} \sum_{\nu_{N-1}} q_{\nu_{N-1}-\nu_N}p_{\nu_{N-1}} \\ \times\ldots\cdot \sum_{\nu_2} q_{\nu_2-\nu_3}p_{\nu_2} \sum_{\nu_1} q_{\nu_1-\nu_2}p_{\nu_1} \;.
\end{multline}
Thus, we may rewrite $\langle Q(\tilde\theta_1,\ldots,\tilde\theta_N) \rangle$ in the form of a trace over the $N$.th power of a matrix $\cal Z$:
\begin{equation}
\label{eq:Theory2.8}
\left\langle Q(\tilde\theta_1,\ldots,\tilde\theta_N) \right\rangle = \sum_{\nu_1} \left[ {\cal Z}^N \right]_{0\nu_1} = \hat e_0^T \cdot {\cal Z}^N \cdot \mathbf n \;,
\end{equation}
where $[\hat e_0]_k = \delta_{0k}$ denotes a unit vector, and $\mathbf n = (\ldots,1,1,1,\ldots)^T$ is composed of unit entries.  The matrix $\cal Z$ itself is defined via:
\begin{equation}
\label{eq:Theory2.9}
{\cal Z}_{jk} = q_{k-j}p_k \;,
\end{equation}
and may be interpreted as the product of a circulant matrix ${\cal Q}_{jk} = q_{k-j}$ with a diagonal matrix ${\cal P}_{jk} = p_j\delta_{jk}$.  (Occasionally, it is helpful to split off a scaling factor $\lambda$ from the coefficients $q_\nu$:  $\tilde q_\nu = \lambda^\nu q_\nu$, e.~g.\ to accomodate functionals in a rotated frame of reference $Q_\alpha(\tilde\theta_1,\ldots,\tilde\theta_N) = Q(\tilde\theta_1 -\alpha,\ldots,\tilde\theta_N -\alpha)$, where $\tilde q_\nu = {\rm e}^{{\rm i}\nu\alpha} q_\nu$ (\ref{eq:Theory2.6}).  A brief inspection of (\ref{eq:Theory2.7}) shows that a slight generalization of (\ref{eq:Theory2.8}) will cover this extension:
\begin{equation}
\label{eq:Theory2.10}
\left\langle Q_\lambda(\tilde\theta_1,\ldots,\tilde\theta_N) \right\rangle = \sum_{\nu_1} \lambda^{\nu_1}\left[ {\cal Z}^N \right]_{0\nu_1} \;.)
\end{equation}

The expressions (\ref{eq:Theory2.8}) and (\ref{eq:Theory2.10}) become especially handy when only a finite number of angular Fourier components $p_k$ in (\ref{eq:Theory0.2}) are non-vanishing.  If $p_\nu\equiv 0$ for all $|\nu|>m$, only the coefficients $q_{-2m},\ldots,q_{2m}$ will enter the calculation (irrespective of the nature of $q(\tilde\theta)$), and the problem in (\ref{eq:Theory2.8}) is effectively reduced to finding the $N.$th power of a $(2m+1)\times(2m+1)$--matrix.  Eigenvalue methods then offer an efficient algorithm to calculate the average for any $N$.  In conclusion, we remark that our matrix formulation of the planar persistent random walk problem closely resembles the treatment of polymer chains in the rotational isomeric state (RIS)--approximation \cite{Flory1969a}, and bears similarity to the Ising model of a one-dimensional ferromagnetic system \cite{Wannier1941a}.  The mathematical aspects of the representation of averages in Markoff chains through powers of a matrix have been summarized by Montroll \cite{Montroll1947a} who incidentally applied the method to discuss the probability distribution function for a persistent random walk in a single dimension.

As an important example for this formalism we examine the characteristic function for the end-to-end probability distribution \cite{Weiss1994a} $c(\omega,\chi) = \langle \exp({\rm i}\boldsymbol\omega\cdot\mathbf L) \rangle_N$ (where $\boldsymbol\omega = (\omega\cos\chi,\omega\sin\chi)^T$).  The projection of the vector $\mathbf L = (L_x,L_y)^T$ onto the axis $\boldsymbol\omega$ is given by $\boldsymbol\omega\cdot\mathbf L = \omega l \sum_k \cos(\tilde\theta_k - \chi)$.  Therefore, the function $q(\tilde\theta)$ (\ref{eq:Theory2.6}) reads in this case:
\begin{equation}
\label{eq:Theory2.11}
q(\tilde\theta) = {\rm e}^{{\rm i}\omega l\cos(\tilde\theta-\chi)} =
\sum_\nu {\rm i}^\nu {\rm e}^{-{\rm i}\nu\chi} J_\nu(\omega l) {\rm e}^{{\rm i}\nu\tilde\theta} \;,
\end{equation}
since $q(\tilde\theta)$ coincides with the generating function for the Bessel functions of integer index \cite{Watson1944a}.  We take advantage of the scaling symmetry (\ref{eq:Theory2.10}) to identify $\lambda = {\rm i}{\rm e}^{-{\rm i}\chi}$ and $q_\nu = J_\nu(\omega l)$, and consequently obtain for the characteristic function:
\begin{equation}
\label{eq:Theory2.12}
c(\omega,\chi) = \sum_\nu {\rm i}^\nu {\rm e}^{-{\rm i}\nu\chi} \left[ {\cal Z}(\omega l)^N \right]_{0\nu} \;,
\end{equation}
where ${\cal Z}(\omega l)_{jk} = J_{k-j}(\omega l) p_k$ (\ref{eq:Theory2.9}).  (In form of a recursive relation, this result is also stated in Ref.~\cite{Larralde1997a}.)  Equation~(\ref{eq:Theory2.12}) may be interpreted as the angular Fourier series for the characteristic function $c(\omega,\chi)$.  Remarkably, if $p_m$ is the non-vanishing Fourier component of largest index $|m|$ in the series for $p(\theta)$ (\ref{eq:Theory0.2}), then also the angular Fourier expansion of $c(\omega,\chi)$ will be a finite series limited to the range $|\nu | \leq |m|$.  In this sense, the angular decomposition of the end-to-end distribution directly reflects the properties of the angular bias $p(\theta)$ between the individual segments that make up the walk.

Finally, we note that our treatment covers Pearson's problem \cite{Pearson1905a} as the isotropic case $p(\theta)=1/2\pi$, or $p_k = \delta_{0k}$ (\ref{eq:Theory0.2}).  Then, the characteristic function (\ref{eq:Theory2.12}) is also isotropic and reduces to a simple power of a Bessel function, $c_P(\omega) = J_0(\omega l)^N$ \cite{Weiss1994a}.

\subsection{The probability distribution function}
\label{sec:Theory3}

It has been pointed out by various authors \cite{Weiss1994a,Flory1969a} that the probability distribution function $w(R,\phi) = \langle \delta[\mathbf R - \mathbf L(\tilde\theta_1,\ldots,\tilde\theta_N)] \rangle_N$ is connected to the characteristic function $c(\omega,\chi) = \langle \exp({\rm i}\boldsymbol\omega\cdot \mathbf L) \rangle_N$ via a Fourier transform in $\boldsymbol\omega$--space:
\begin{equation}
\label{eq:Theory3.1}
w(R,\phi) = \frac1{(2\pi)^2}\, \int {\rm d}^2\boldsymbol\omega\, {\rm e}^{-{\rm i}\boldsymbol\omega \cdot \mathbf R}\, c(\omega,\chi) \;.
\end{equation}
Since $\boldsymbol\omega\cdot\mathbf R = \omega R\cos(\chi-\phi)$, we may expand the exponential in (\ref{eq:Theory3.1}) into the series of Bessel functions (\ref{eq:Theory2.11}).  Inserting the matrix-type expression for $c(\omega,\chi)$ (\ref{eq:Theory2.12}) renders the integration over the polar angle $\chi$ in (\ref{eq:Theory3.1}) trivial, and one obtains an angular Fourier series for the end-to-end distribution function:
\begin{equation}
\label{eq:Theory3.2}
w(R,\phi) = \frac1{2\pi} \sum_\nu {\rm e}^{-{\rm i}\nu\phi}
\int_0^\infty \negmedspace \omega{\rm d}\omega \,
J_\nu(\omega R) \left[ {\cal Z}(\omega l)^N \right]_{0\nu}\;.
\end{equation}
We may interpret this result as the generalization of Kluyver's formula \cite{Kluyver1906a} towards the persistent random walk, to which it reduces in the isotropic case $p_k=\delta_{0k}$:
\begin{equation}
\label{eq:Theory3.3}
w_P(R) = \frac1{2\pi} \int_0^\infty \negmedspace \omega{\rm d}\omega \,
J_0(\omega R) J_0(\omega l)^N \;.
\end{equation}
As for $c(\omega,\chi)$, we again infer that the range of non-vanishing angular Fourier components in $w(R,\phi)$ equals the corresponding range for the local weight function $p(\theta)$ concerning the individual angles between steps.

While a rather elegant expression, (\ref{eq:Theory3.2}) is nevertheless not well suited towards the accurate determination of $w(R,\phi)$, since this pursuit involves the numerical evaluation of an oscillating integrand (a sum over products of $N+1$ individual Bessel functions, see (\ref{eq:Theory2.7})).  Indeed, since $w(R,\phi)=0$ for $R>Nl$ (a fact that we will take advantage below), the distribution is not an analytic function of $\mathbf R$, and the asymptotics of the integrand in (\ref{eq:Theory3.2}) suggests that $w(R,\phi)$ shows singular behavior whenever the lengths $\pm R$, $\pm l$ involved sum up to zero \cite{Flory1969a}.  (For identical step lengths $l$, this occurs at regular intervals $R = (N-2k)l$, where $k=0,1,2,\ldots,N$.)  From the practical point of view, the eigenfunction expansion of $w(R,\phi)$ that we are going to develop now is much superior.  On the other hand, the analytical expressions for the integral (\ref{eq:Theory3.3}) known for $N\leq 3$ provide a useful testing ground for any numerical attempt to evaluate the probability distribution function (Section~\ref{sec:Example}).

\subsection{The eigenfunction method}
\label{sec:Theory4}

Since the end-to-end distance $R$ of a walk of $N$ steps, regardless of its details, cannot exceed the contour length $Nl$ of the chain, the distribution function $w(R,\phi)$ vanishes outside the disc $R \leq Nl$.  This renders a technique attractive that is familiar from the theory of field equations under boundary conditions, viz., the expansion into eigenfunctions of a self-adjoint operator.  For a function confined within a circle, the modes of a drum, i.~e.\ solutions of the equation $(\Delta + \lambda^2_{mk}) f_{mk}(R,\phi) = 0$, provide a natural choice.  Properly normalized, these functions read \cite{Sommerfeld1949a}:
\begin{equation}
\label{eq:Theory4.1}
f_{mk}(R,\phi) = \frac{{\rm e}^{-{\rm i}m\phi}J_m(z_{mk}R/Nl)}{Nl \sqrt\pi\, J_{m+1}(z_{mk})} \;,
\end{equation}
where $J_m(z)$ is the Bessel function of order $m$, $m\in{\mathbb Z}$, $z_{mk}$ denotes the $k$.th zero of $J_m(z)$ ($k\geq 1$), and $\lambda_{mk} = z_{mk}/Nl$ is the assigned eigenvalue.  The functions (\ref{eq:Theory4.1}) form an orthonormal set that is furthermore complete, i.~e., with $L_x = L \cos\psi$ we find for all $R,L < Nl$:
\begin{equation}
\label{eq:Theory4.2}
\delta(\mathbf R - \mathbf L) = \sum_{m=-\infty}^\infty \sum_{k=1}^\infty f_{mk}(L,\psi)^* f_{mk}(R,\phi) \;.
\end{equation}
Now, the probability distribution function $w(R,\phi) = \langle \delta[\mathbf R - \mathbf L(\tilde\theta_1,\ldots,\tilde\theta_N)] \rangle_N$ presents just the average of (\ref{eq:Theory4.2}), and we thus formally obtain \cite{Weiss1987a}:
\begin{equation}
\label{eq:Theory4.3}
w(R,\phi) = \sum_{m=-\infty}^\infty \sum_{k=1}^\infty \frac{F_{mk} {\rm e}^{-{\rm i}m\phi} J_m\left(z_{mk}R/Nl \right)}{(Nl)^2 \pi J_{m+1}(z_{mk})^2}
\end{equation}
as the Fourier-Bessel expansion of $w(R,\phi)$, where the weight coefficients $F_{mk}$ are given by the averages:
\begin{equation}
\label{eq:Theory4.4}
F_{mk} = \left\langle {\rm e}^{{\rm i}m\psi(\tilde\theta_1,\ldots,\tilde\theta_N)} J_m\left( \frac{z_{mk}}{Nl}\, L(\tilde\theta_1,\ldots,\tilde\theta_N) \right) \right\rangle_N \;.
\end{equation}

The actual calculation of the coefficients $F_{mk}$ follows the general line of argument presented in Section~\ref{sec:Theory2}, where only a slight modification is required.  The weights $F_{mk}$ present averages of functionals of the absolute angles of the segments $\tilde\theta_\nu$ in the sense of (\ref{eq:Theory2.2}), and we thus seek to establish the angular Fourier series of $Q_{mk}(\tilde\theta_1,\ldots,\tilde\theta_N) = {\rm e}^{{\rm i}m\psi} J_m\left( z_{mk}L/Nl\right)$.  Although this may appear as a formidable task at first, an easily proven ``projection theorem'' (\ref{eq:Proj1}) for Bessel functions (Appendix~\ref{sec:Projection}) will provide the desired result:
\begin{equation}
\label{eq:Theory4.5}
Q^{(mk)}_{\mu_1,\ldots,\mu_N} = \delta_{m,\mu_1+\ldots+\mu_N} J_{\mu_1}(z_{mk}/N) \cdot\ldots\cdot J_{\mu_N}(z_{mk}/N) \;.
\end{equation}
Inserting this result into (\ref{eq:Theory2.4}) and switching to the indexing scheme $\{\nu_1,\ldots,\nu_N\}$ (\ref{eq:Theory2.5}), we find that $\nu_1 = m$.  Rearranging the remaining summations in the spirit of (\ref{eq:Theory2.7}) yields at once:
\begin{multline}
\label{eq:Theory4.6}
F_{mk} = \sum_{\nu_N} J_{\nu_N}\!\left( \frac{z_{mk}}N \right) p_{\nu_N}
\sum_{\nu_{N-1}} J_{\nu_{N-1} - \nu_N}\!\left( \frac{z_{mk}}N \right) p_{\nu_{N-1}} \\
\times\ldots\cdot \sum_{\nu_2} J_{\nu_2-\nu_3}\left(\frac{z_{mk}}N \right) p_{\nu_2} J_{m - \nu_2}\left(\frac{z_{mk}}N \right) p_m \;.
\end{multline}
By comparison with (\ref{eq:Theory2.8}), we infer that this multiple sum can be recast again into the $N$.th power of a matrix ${\cal Z}(z_{mk}/N)$: $F_{mk} = \left[{\cal Z}(z_{mk}/N)^N\right]_{0m}$, whose matrix elements are given by:
\begin{equation}
\label{eq:Theory4.7}
\left[{\cal Z}(z_{mk}/N)\right]_{\mu\nu} = J_{\nu-\mu}\left( \frac{z_{mk}}N \right) p_\nu \;.
\end{equation}
In conjunction with (\ref{eq:Theory4.3}), this finally yields the Fourier--Bessel expansion of the end-to-end distribution function $w(R,\phi)$ of the persistent random walk:
\begin{multline}
\label{eq:Theory4.8}
w(R,\phi) = \sum_{m=-\infty}^\infty \sum_{k=1}^\infty \frac{ \left[{\cal Z}(z_{mk}/N)^N\right]_{0m} }{(Nl)^2 \pi J_{m+1}(z_{mk})^2} \\ \times\,{\rm e}^{-{\rm i}m\phi} J_m\left( \frac{z_{mk}R}{Nl} \right) \;.
\end{multline}
Several remarks are in order here.  First, we point out that the expansion coefficient $F_{mk}$ (\ref{eq:Theory4.6}) vanishes if the corresponding internal angular Fourier component $p_m$ does.  Thus, the  angular character of the end-to-end distribution function $w(R,\phi)$ and the angular bias $p(\theta)$ at the joints of individual segments (\ref{eq:Theory0.2}) are directly related.  Like in Section~\ref{sec:Theory2}, a terminating Fourier expansion of $p(\theta)$ implies a finite-size matrix ${{\cal Z}(z_{mk}/N)}$ in the determination of the coefficients in (\ref{eq:Theory4.8}).  In particular, for Pearson's isotropic walk the radial distribution function $w_P(R)$ (\ref{eq:Theory3.3}) simplifies to:
\begin{equation}
\label{eq:Theory4.9}
w_P(R) = \sum_{k=1}^\infty \frac{ J_0(z_{0k}/N)^N }{(Nl)^2 \pi J_1(z_{0k})^2} \, J_0\left( \frac{z_{0k}R}{Nl} \right) \;,
\end{equation}
a result first obtained by Barakat \cite{Barakat1974a}.

Second, series similar in structure to (\ref{eq:Theory4.8}) account for arbitrary derivatives of the probability distribution function $w(R,\phi)$.  We invoke the following differentiation relation \cite{Watson1944a} for the eigenfunctions $f_{mk}(R,\phi)$ (\ref{eq:Theory4.1}):
\begin{multline}
\label{eq:Theory4.10}
\left( \frac\partial{\partial x} \pm {\rm i}\,\frac\partial{\partial y} \right)^n
\left[ {\rm e}^{-{\rm i}m\phi} J_m\left( \frac{z_{mk}R}{Nl} \right) \right] = \\
\left( \pm \frac{z_{mk}}{Nl} \right)^n
\left[ {\rm e}^{-{\rm i}(m \mp n)\phi} J_{m\mp n}\left( \frac{z_{mk}R}{Nl} \right) \right] \;,
\end{multline}
(which immediately yields the eigenfunction property $\Delta f_{mk}(R,\phi) = -\lambda_{mk}^2 f_{mk}(R,\phi)$  of these solutions to the wave equation) and apply it to (\ref{eq:Theory4.8}):
\begin{multline}
\label{eq:Theory4.11}
\left(\partial_x + {\rm i}\partial_y \right)^r \left(\partial_x -{\rm i} \partial_y \right)^n w(R,\phi) = (-1)^n \\ \times \sum_{m=-\infty}^\infty \sum_{k=1}^\infty
\left( \frac{z_{mk}}{Nl} \right)^{n+r}  \frac{ \left[{\cal Z}(z_{mk}/N)^N\right]_{0m} }{(Nl)^2 \pi J_{m+1}(z_{mk})^2} \\ \times\,{\rm e}^{-{\rm i}(m+n-r)\phi}  J_{m+n-r}\left( \frac{z_{mk}R}{Nl} \right) \;.
\end{multline}
All derivatives of the form $\partial_x^\mu \partial_y^\nu w(R,\phi)$ are then available as properly chosen linear combinations of the Fourier--Bessel series (\ref{eq:Theory4.11}), where $n+r = \mu+\nu$.  The representation (\ref{eq:Theory4.11}) is particularly useful in the random walk model of polymers \cite{Flory1969a}, where the gradient $\bm\nabla\log w(R,\phi)$ is related to the average force exerted by the polymer chain.

Finally, also all physically sensible moments $\langle x^\mu y^\nu \rangle_N$ ($\mu,\nu \in {\mathbb N}_0$) of the probability distribution function in the persistent walk can be represented as fairly simple sums involving the zeroes $z_{mk}$ of Bessel functions (\ref{eq:Mom5}), (\ref{eq:Mom6}).  Since the evaluation of these moments involves an integral over Bessel functions that is not contained in standard tables \cite{Gradshteyn1980a}, the details of the calculation have been assembled in Appendix~\ref{sec:Moments}.

\section{Example calculations}
\label{sec:Example}

In the following, we probe different practical aspects of the Fourier--Bessel expansion (\ref{eq:Theory4.8}) derived in the previous section.  The numerical convergence of the method is the first topic to be examined in detail, and we present analytical and numerical results obtained on Pearson's original walk problem below (Section~\ref{sec:Example1}).  Some of the distinctive properties of persistent walks are highlighted in Section~\ref{sec:Example2} where we will study how the underlying angular bias $p(\theta)$ influences the shape of the probability distribution function.  Our examples include a ``chiral'' walk problem \cite{Larralde1997a} that is biased towards left or right turns, and a model illustrating the transition from diffusive to elastic behavior as the stiffness of the chain is increased.

\subsection{The Pearson random walk}
\label{sec:Example1}

Obviously, particularly simple results are obtained when the Fourier--Bessel method of Section~\ref{sec:Theory4} is applied towards the isotropic Pearson random walk with $p(\theta) = 1/2\pi$ or $p_\nu = \delta_{0\nu}$ (\ref{eq:Theory0.2}).  The angular sum in (\ref{eq:Theory4.8}) is then absent, and the matrix power in the Fourier coefficient reduces to a simple power, which allows convenient access to error estimates for the truncated sums necessarily encountered in a numerical evaluation of $w_P(R)$ via the expansion (\ref{eq:Theory4.9}).

\begin{figure}
\centerline{\includegraphics[draft=false,width=\columnwidth]{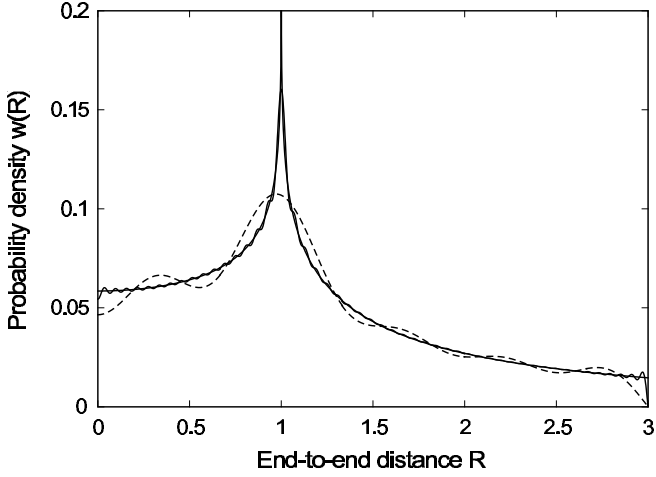}}
\caption{The probability distribution function $w_P(R)$ for a three-step Pearson walk.  The exact solution (\ref{eq:Example1.1})--(\ref{eq:Example1.2}) (bold line) is shown together with the Fourier--Bessel expansion (\ref{eq:Theory4.9}), truncated after $k=10$ terms (dashed line) and $k = 100$ terms (solid line).  For $k>1000$, the approximation becomes visually indistinguishable from the exact solution.
\label{fig:n3pdf}}
\end{figure}

Unfortunately, the probability distribution function $w_P(R)$ for a random walk in two dimensions, unlike its three-dimensional counterpart \cite{Flory1969a,Treloar1946a}, generally defies evaluation in closed form.  However, analytical expressions can be found for the shortest chains with $N=2$ and $N=3$.  A formula for the two-segment chain distribution is available for arbitrary persistent walks (see Section~\ref{sec:Example2}), and reads here $w_P(R) = [\pi^2 R \sqrt{4l^2-R^2}]^{-1}$ (\ref{eq:Example2.1}).  For $N=3$, the integral representation (\ref{eq:Theory3.3}) can be evaluated in closed form \cite{Watson1944a}.  Denoting the complete elliptic integral of the first kind \cite{Gradshteyn1980a} by $K(k)$, we find:
\begin{multline}
\label{eq:Example1.1}
w_P(R) = \frac2{\pi^3 (l+R) \sqrt{(3l-R)(l+R)}} \\
\times\, K\left[ \frac{4l}{l+R} \sqrt{\frac{lR}{(3l-R)(l+R)}} \right] \;,
\end{multline}
in the interval $0 \leq R < l$, and
\begin{equation}
\label{eq:Example1.2}
w_P(R) = \frac1{2\pi^3 l \sqrt{lR}} \, K\left[ \frac{l+R}{4l} \sqrt{\frac{(3l-R)(l+R)}{lR}} \right] \;,
\end{equation}
for $l < R \leq 3l$.  Note that $w_P(R)$ possesses a logarithmic singularity at $R = l$ and a steplike discontinuity at the maximum extension $R=3l$, where it approaches the value $w_P(3l) = 1/(4\sqrt3 \pi^2l^2)$.  (The division of $w_P(R)$ into distinct intervals of length $2l$ was already mentioned at the end of Section~\ref{sec:Theory3}.)  Its exact shape is displayed in Figure~\ref{fig:n3pdf}.

\begin{figure}
\centerline{\includegraphics[draft=false,width=\columnwidth]{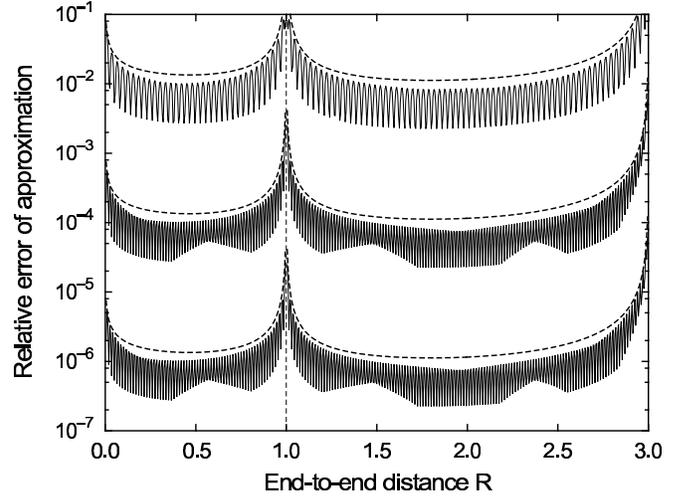}}
\caption{Relative error $\epsilon_{\mathrm{rel}}(k,R)$ incurred in the truncation of the Fourier-Bessel series (\ref{eq:Theory4.9}) for $N=3$.  From top to bottom, the solid lines display the relative error of the series for $k=100$, $k=10^4$, and $k=10^6$, as compared to the exact result (\ref{eq:Example1.1}) and (\ref{eq:Example1.2}).  The heuristic error estimate (\ref{eq:Example1.5}) (dashed lines) shows that the error scales inversely with $k$.
\label{fig:n3error}}
\end{figure}

These formulas provide a useful testing ground for the Fourier--Bessel expansion (\ref{eq:Theory4.9}).  Clearly, this series cannot be convergent in a strict sense since $w_P(R)$ diverges as $R\rightarrow l$.  For an approximate analysis, we estimate the asymptotic value of the sum terms $S_\mu(N,R)$ with large index $\mu$.  Since $z_{0\mu} \sim \pi(\mu - \frac14)$ and $J_0(\lambda z_{0\mu}) \sim \sqrt{2/\pi\lambda z_{0\mu}}\cos(\lambda z_{0\mu} - \pi/4)$ holds as $\mu\rightarrow \infty$ \cite{Watson1944a}, we find for $R \gg Nl/\mu$:
\begin{multline}
\label{eq:Example1.6}
S_\mu(N,R) \sim \frac1{\pi N l^2} \left( \frac{2N}{\pi^2 \mu} \right)^{\frac{N-1}2} \sqrt{\frac lR} \\ \times\,
\cos\left[ \frac{\pi(4\mu-N-1)}{4N} \right]^N \cos\left[ \frac{\pi R(4\mu - 1)}{4Nl} -\frac\pi4 \right] \;.
\end{multline}
Thus, the series terms overall decay like $S_\mu(N,R) \sim \mu^{-(N-1)/2}$.  Equation~(\ref{eq:Example1.6}) allows us to examine the absolute error $\epsilon(k,R)$ introduced by truncating the series (\ref{eq:Theory4.9}) to terms with index $\mu \leq k$.  This quantity is the sum of the neglected terms, $\epsilon(k,R) = \sum_{\mu > k} S_\mu(N,R)$, and its asymptotics depend on the behavior of the oscillating terms in the second line of (\ref{eq:Example1.6}).  If $N - R/l$ is not en even integer, the average of the oscillating term vanishes, the series is of the alternating type, and for an estimate of $\epsilon(k,R)$ we may multiply $S_k(N,R)$ with the largest period present in the product of cosine functions.  These periods turn out to be  proportional to the inverse distance of $R$ to the interval boundaries $(N - 2\nu)l$ (see Section~\ref{sec:Theory3}).  Indeed, for $N=3$ the maximum error $\epsilon(k,R)$ incurred in the truncation of the series is approximated well by the empiric formula:
\begin{equation}
\label{eq:Example1.5}
\epsilon(k,R) \approx \frac{0.0076}{kl\sqrt{lR}} \left( \frac12 + \frac{3l}{|l-R|} +\frac{2l}{3l-R} \right) \;.
\end{equation}
In the case that $Nl-R$ is an even multiple of $l$, the Fourier series contains a strictly positive (non-oscillating) part that dominates the remainder.  Its summation yields the estimate $\epsilon(k,R) \sim k^{-(N-3)/2}$ for the interval boundaries, except for $N=3$, where it reproduces the logarithmic singularity at $R=l$ (\ref{eq:Example1.1}).  Examples for the numerical convergence of the Bessel--Fourier series for $N=3$ are shown in Figure~\ref{fig:n3pdf} and Figure~\ref{fig:n3error}, where the relative error $\epsilon_{\mathrm{rel}}(k,R) = \epsilon(k,R)/w_P(R)$ is displayed together with the estimate (\ref{eq:Example1.5}) for various values of $k$.

\begin{figure}
\centerline{\includegraphics[draft=false,width=\columnwidth]{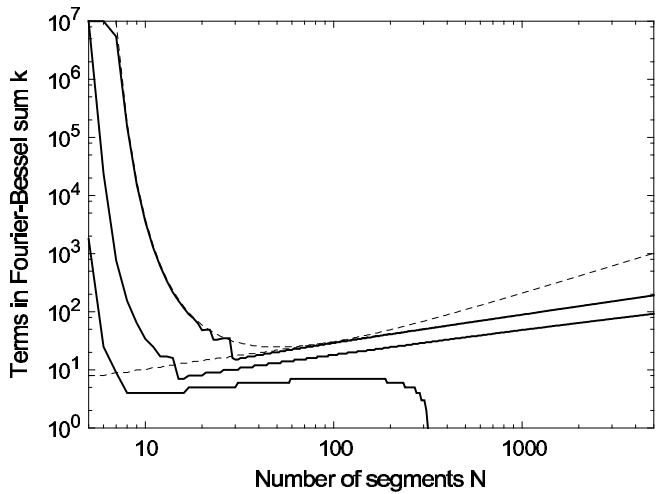}}
\caption{Number of terms $k$ in the Fourier-Bessel sum (\ref{eq:Theory4.9}) required to guarantee a prescribed absolute accuracy $\epsilon$ in $w_P(R)$.  Solid lines, from top to bottom: $\epsilon =10^{-12}$, $10^{-6}$, $10 ^{-3}$.  The dotted lines display the short-chain and long-chain approximations (\ref{eq:Example1.3}), (\ref{eq:Example1.4}) for $\epsilon =10^{-12}$, respectively.  (Note that for $N>316$, $w_P(R) < 10^{-3}$ holds. For $\epsilon = 10^{-12}$ and $N=5,6$, numerical convergence has not been achieved.)
\label{fig:fourier}}
\end{figure}

Clearly, for fixed $N$ the absolute convergence of the expansion (\ref{eq:Theory4.9}) is slowest at the origin $R=0$.  We may then use arguments similar to (\ref{eq:Example1.6}) to establish an absolute error bound $\epsilon(k)$ for the Fourier--Bessel series, which then yields $k$ as a function of the desired accuracy $\epsilon$:
\begin{equation}
\label{eq:Example1.3}
k(\epsilon) \sim \frac{4N}{\pi^2} \left[ \frac{16 \Gamma\left( \frac{N+2}4 \right)^2}{\pi^4 (N-4) \Gamma\left( \frac{N+2}2 \right) \epsilon} \right]^{\frac2{N-4}} \;,
\end{equation}
Note that for short chains, the absolute series convergence at $R=0$ ($k \sim \epsilon^{-2/(N-4)}$) is much worse than its typical convergence ($k \sim \epsilon^{-2/(N-1)}$).  In particular, (\ref{eq:Example1.3}) predicts the divergence of $w_P(R)$ for $N=4$ as $R$ approaches the origin.  Invoking the general relation $w_P(N+1,R=0) = w_P(N,R=l)$ that follows at once from (\ref{eq:Theory3.3}), the analytic expressions (\ref{eq:Example1.1}), (\ref{eq:Example1.2}) immediately confirm this assertion.  With increasing number of segments $N$, the number of sum terms $k(\epsilon)$ drops rapidly, and finally becomes of the order of the chain length itself, in which case the estimate (\ref{eq:Example1.3}) becomes too conservative.  In the long-chain regime $\pi k(\epsilon) \ll N$ the argument of the Bessel function in (\ref{eq:Theory4.9}) tends towards zero, and the approximation $J_0(z_{0k}/N)^N \sim (1- z_{0k}^2/4N^2)^N \sim \exp(- z_{0k}^2/4N)$ instantly yields the asymptotic behavior:
\begin{equation}
\label{eq:Example1.4}
k(\epsilon) \sim \frac2\pi \sqrt{-N \log(\pi N\epsilon)} \;.
\end{equation}
Therefore, for long chains, the numerical complexity of the Fourier--Bessel series approach grows merely with $(N\log N)^{1/2}$.  Figure~\ref{fig:fourier} depicts numerical results for $k(\epsilon)$ for various values of the absolute accuracy $\epsilon$, together with the estimates (\ref{eq:Example1.3}) and (\ref{eq:Example1.4}).  We conclude that these expressions closely reproduce $k(\epsilon)$ within their respective regimes of validity.

\begin{figure}
\centerline{\includegraphics[draft=false,width=\columnwidth]{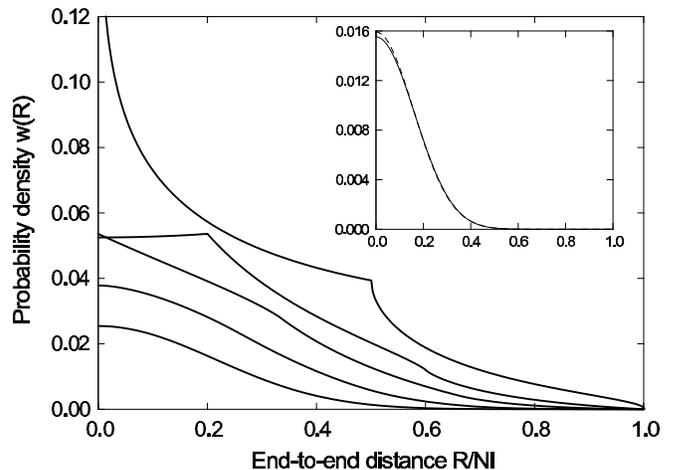}}
\caption{Dependence of the probability distribution function $w_P(R)$ on the number of steps $N$.  From top to bottom, the distributions for $N=4,5,6,8$, and $N=12$ are displayed.  The nonanalytic behavior of $w_P(R)$ at the interval boundaries $R = (N-2\nu)l$ is clearly distinguishable for short chains.  (For $N=4$, $w_P(R)$ diverges logarithmically as $R\rightarrow 0$ (\ref{eq:Example1.3}).)   As $N$ increases, the diffusive character prevails.  The inset shows the distribution for $N=20$ (solid line), together with the Gaussian approximation (\ref{eq:Example1.7}) (dotted line).
\label{fig:pdf}}
\end{figure}

Following this excursion on the numerical aspects of the Fourier-Bessel approach, we next study the resulting probability distributions $w_P(R)$ (\ref{eq:Theory4.9}) for the Pearson random walk.  As a random process, $w_P(R)$ must be normally distributed in the limit $N\rightarrow \infty$, and thus is completely determined by the diffusion constant $D_P = l^2$ (\ref{eq:Theory1.7}).  After proper normalization, $w_P(R)$ asymptotically behaves like:
\begin{equation}
\label{eq:Example1.7}
w_P(R) \sim \frac1{\pi Nl^2}\exp\left( - \frac{R^2}{Nl^2}\right) \;,
\end{equation}
assuming that $R \ll Nl$.  For short walks of less than $N=10$ segments, deviations from the Gaussian limit become clearly apparent, and the separation of $w_P(R)$ into distinct functions for every interval of length $2l$ reveals itself.  The nonanalytic behavior of $w_P(R)$ at the interval boundaries $R = (N-2\nu)l$ is manifested by the presence of divergences ($N=3,4$) and kinks ($N=4,5,6$) in the distribution functions, as shown in Figure~\ref{fig:pdf}.

\begin{figure}
\centerline{\includegraphics[draft=false,width=\columnwidth]{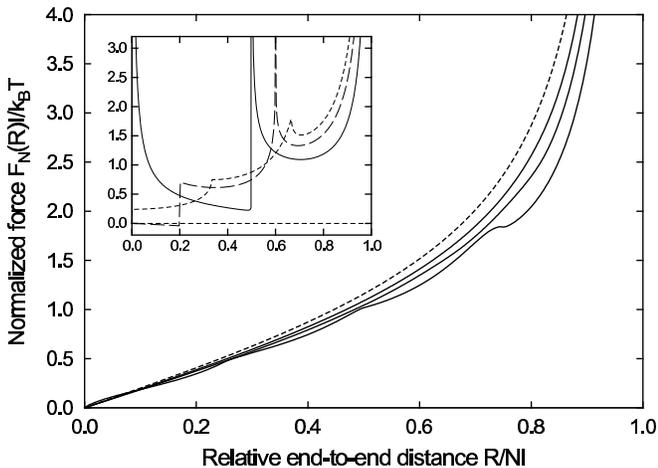}}
\caption{Average force $F_N(R)$ exerted by a freely rotating chain as a function of the end-to-end distance $R$.  The plot shows the logarithmic derivative of $w_P(R)$ for various chain lengths $N$. In the diffusive limit $N\rightarrow \infty$, the force curves (solid lines, bottom to top: $N=8,12,20$) tend towards the function $F_\infty(R)$ (\ref{eq:Example1.8}) (dotted line).  The inset displays $F_N(R)$ for short chains (solid line: $N=4$, dashed line: $N=5$, dotted line: $N=6$).  The peculiar appearance of these functions is linked to the nonanalytic behavior of $w_P(R)$ (Figure~\ref{fig:pdf}).
\label{fig:force}}
\end{figure}

Finally, we examine the force function $F_N(R)$ assigned to the random walk problem.  Persistent random walks have been proposed as a model for the configuration of polymer chains \cite{Flory1969a}.  In this statistical interpretation, the bias function $p(\theta)$ (\ref{eq:Theory0.2}) is related to the angular potential between adjacent segments $U(\theta)$ via (properly normalized) Boltzmann factors $p(\theta) \propto \exp[-U(\theta)/k_BT]$ which renders the Pearson walk equivalent to the uncorrelated freely joined chain (FJC) model of the polymer.  For a fixed end position $\mathbf R$ of the polymer chain, the molecule exerts an entropic, temperature-dependent force whose average is given by the logarithmic derivative of the distribution function \cite{Flory1969a}: $\langle \mathbf F(\mathbf R) \rangle = k_BT  \,\bm\nabla \log w(\mathbf R)$, and its absolute value reduces to $F_N(R) = -k_BT w_P(R)'/w_P(R)$ for the isotropic walk.  Like the distribution function itself, its derivatives are conveniently calculated using the Fourier--Bessel formalism (\ref{eq:Theory4.11}), and the results of this effort are displayed in Figure~\ref{fig:force} for the two-dimensional FJC model.

For short chains, the singularities of $w_P(R)$ at the interval boundaries $R=(N-2\nu)l$ confer a conspicious irregular shape to the corresponding force curves $F_N(R)$.  As $N$ increases, $F_N(R)$ becomes smoother, and as a function of the scaled end-to-end distance $R/Nl$, approaches a well-defined limit $F_\infty(R)$.  Its analytic form may be obtained by considering the FJC in the Gibbs ensemble, where the force $\mathbf F$ is kept constant, and $\mathbf R$ is allowed to vary \cite{Flory1969a}.  Its partition function is given by $Z(\mathbf F) = \langle \exp(-\mathbf L\cdot\mathbf F / k_BT) \rangle$, and the average position of the chain end $\langle \mathbf R \rangle$ under the applied force $\mathbf F$ follows at once from $\langle \mathbf R(\mathbf F) \rangle_N = - k_BT \,\bm\nabla \log Z(\mathbf F)$.  Using the technique presented in Section~\ref{sec:Theory2}, the function $Z(\mathbf F)$ can be cast as the trace over a matrix power (\ref{eq:Theory2.8}) for an arbitrary persistent walk.  Indeed, $Z(\mathbf F)$ may be interpreted as the characteristic function $c(\bm\omega)$ for imaginary argument $\bm\omega = {\rm i}\mathbf F/k_BT$ (\ref{eq:Theory2.12}), and thus reduces for the isotropic walk to $Z_P(F) = I_0(lF/k_BT)^N$, where $I_0(z)$ denotes the modified Bessel function \cite{Watson1944a}.  Hence, for the FJC model we find $\langle R(F) \rangle_N = Nl I_1(u)/I_0(u)$, where $u =lF/k_BT$.  In the limit $N\rightarrow \infty$, we may identify the average position $\langle R(F) \rangle_N$ with the actual position $R$, and likewise $F_N(R)$ with $F$.  This allows to revert the relation to express $F_\infty(R)$ as a function of $R/Nl$:
\begin{equation}
\label{eq:Example1.8}
F_\infty(R) = \frac{k_BT}l \, {\cal U}(R/Nl) \;,
\end{equation}
where ${\cal U}(z)$ is the inverse function to $I_1(z)/I_0(z)$.  It is defined for $|z|<1$ and behaves like ${\cal U}(z) = 2z+z^3+\frac56 z^5+\ldots$ for small $|z|$, while its Laurent series near $z=1$ is given by ${\cal U}(1-z) = \frac1{2z} + \frac14 +\frac38 z + \ldots$

\subsection{Properties of persistent walks}
\label{sec:Example2}

Analytical methods towards the calculation of the probability distribution function $w(R,\phi)$ for the more general persistent walk problem are usually limited to the combination of only two segments ($N=2$).  In this case, for a fixed endpoint $\mathbf L = (R\cos\phi, R\sin\phi)$ only two possible locations for the intermediate joint exist, and simple geometrical arguments lead to the following result:
\begin{multline}
\label{eq:Example2.1}
w(R,\phi) = \frac2{R\sqrt{4l^2-R^2}} \\ \times\,
\left[ p(2\gamma)p(\phi-\gamma) + p(-2\gamma)p(\phi+\gamma) \right] \;,
\end{multline}
where we used $\gamma = \arccos(R/2l)$ to denote the angle between the end-to-end vector and the individual segments.  (For the uncorrelated Pearson walk, the term in brackets reduces to $1/(2\pi^2)$.)  Likewise, the limit of very extended walks ($N\rightarrow \infty$) is amenable to an analytical treatment.  Here, the end-to-end probability function $w(R,\phi)$ must be again normally distributed, and is completely determined by the persistence lengths $\langle L_x \rangle_\infty$ (\ref{eq:Theory1.3a}), $\langle L_y \rangle_\infty$ (\ref{eq:Theory1.3b}), and the diffusion constant $D_\infty$ (\ref{eq:Theory1.7}):
\begin{equation}
\label{eq:Example2.2}
w(\mathbf R) \sim \frac1{\pi ND_\infty} \exp\left[ -\,\frac{(\mathbf R - \langle \mathbf L \rangle_\infty)^2}{ND_\infty} \right] \;.
\end{equation}
We point out that the distribution $w(R,\phi)$ for sufficiently large $N$ depends merely on the single angular Fourier component $p_1$.  Little can be said \textsl{a priori} about the behavior of $w(R,\phi)$ for intermediate chain lengths $N$.  Fortunately, the Fourier--Bessel expansion (\ref{eq:Theory4.8}) provides an efficient means to determine the probability distribution function numerically.
\begin{figure}
\centerline{\includegraphics[draft=false,width=0.6\columnwidth]{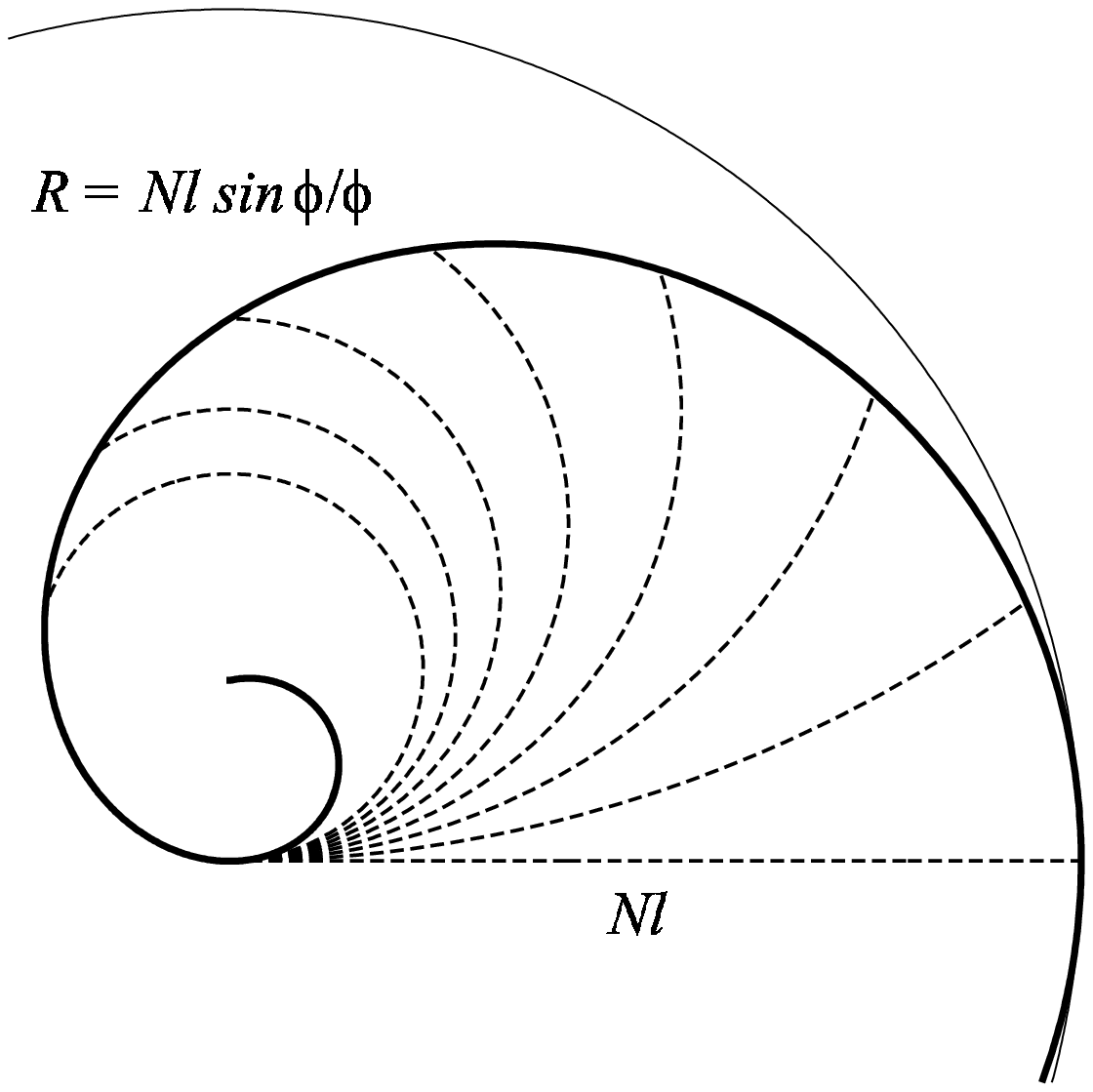}}
\caption{The elastic model of a stiff persistent random walk.  The chains are replaced by circular arcs of fixed contour length $Nl$.  When the curvature of the arc is changed, its end point moves on the geometrical curve $R = Nl\sin\phi/\phi$.
\label{fig:Spring}}
\end{figure}

In the following, we select normalized angular bias functions $p(\theta)$ of the general form:
\begin{equation}
\label{eq:Example2.3}
p(\theta) = \frac{2^{2\xi-1}\Gamma(\xi+1)^2}{\pi\Gamma(2\xi+1)} \cos^{2\xi}\left(\frac{\theta-\beta}2 \right) \;,
\end{equation}
where $\xi\geq 0$ and $\beta$ are real parameters.  The corresponding angular Fourier components $p_\nu$ (\ref{eq:Theory0.2}) are given by \cite{Gradshteyn1980a}:
\begin{equation}
\label{eq:Example2.4}
p_\nu(\xi,\beta) = \frac{\Gamma(\xi+1)^2 {\rm e}^{{\rm i}\nu\beta}}{\Gamma(\xi+\nu+1)\Gamma(\xi-\nu+1)} \;.
\end{equation}
In practice, the coefficients $p_\nu(\xi,\beta)$ are best calculated by the simple recursion formula $p_{\nu+1} = {\rm e}^{{\rm i}\beta}(\xi-\nu)p_\nu / (\xi+\nu+1)$ ($\nu \geq 0$), where normalization requires $p_0=1$.  We find in particular $p_1 = {\rm e}^{{\rm i}\beta}\xi /(\xi+1)$ and thus:
\begin{align}
\label{eq:Example2.5}
\left\langle \mathbf L \right\rangle_\infty & = \frac{l\xi}{1+4\xi(\xi+1)\sin^2\frac\beta2}\left(
\begin{matrix}
1 -2(\xi+1)\sin^2\frac\beta2 \\ (\xi+1)\sin\beta
\end{matrix} \right) \;,
\\
\label{eq:Example2.6}
D_\infty &= \frac{l^2(2\xi+1)}{1+4\xi(\xi+1)\sin^2\frac\beta2} \;,
\end{align}
in (\ref{eq:Example2.2}).  Note that the angle $\beta$ in (\ref{eq:Example2.3}) controls the chiral bias of the walk \cite{Larralde1997a}; a symmetrical distribution will ensue if $\beta = 0$ is chosen.  (Then, (\ref{eq:Example2.5}) simplifies to $\langle L_x \rangle_\infty = l\xi$, $\langle L_y \rangle_\infty = 0$, and $D_\infty = l^2(2\xi+1)$.)  Finally, we remark that the series in (\ref{eq:Example2.3}) terminates whenever $\xi$ is an integer.  (Otherwise, $p(\theta)$ is not analytic at its zero $\theta_0 =\pi+\beta$.)
\begin{figure*}
\centerline{\includegraphics[draft=false,width=\textwidth]{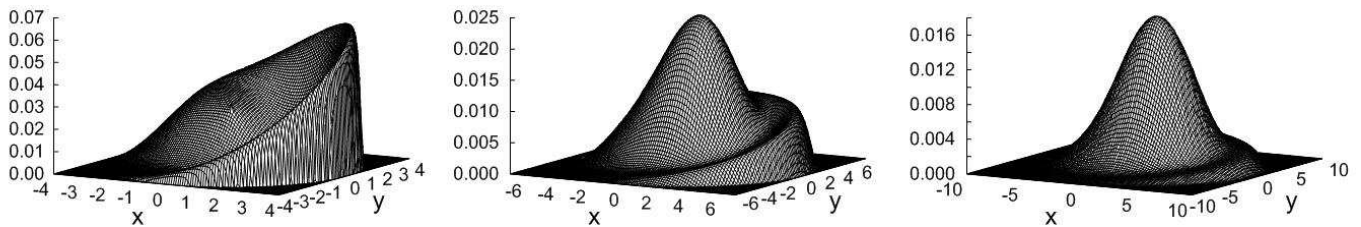}}
\caption{Probability density distribution $w(R,\phi)$ for a persistent random walk with $p(\theta) = \frac4{3\pi}\cos^4[\frac12(\theta - 30^\circ)]$, made up of $N=4$ (left), $N=7$ (center), and $N=10$ (right) segments.  The transition from a stiff chain with a narrow distribution to a diffusive structure with Gaussian character is obvious.  In the intermediate regime, elements of both limiting cases coexist.
\label{fig:Chiral}}
\end{figure*}

We start out with a study of the dependence of $w(R,\phi)$ on the number of steps $N$ for the fixed angular weight $p(\theta) = \frac4{3\pi}\cos^4[\frac12 (\theta - \frac\pi6)]$, which is of the form (\ref{eq:Example2.3}) with $\xi=2$ and $\beta=30^\circ$.  The angular bias evidently favors a ``spiral'' configuration of the chain in which adjacent segments are joined at an angle of $\beta$, which translates to an absolute angle $\phi = (N+1)\beta/2$ of the end-to-end vector with respect to the $x$--axis.  However, configurations that end up near the origin constitute the vast majority in the phase space $\{\theta_1,\ldots,\theta_N\}$ of the walk.  For sufficiently large $N$, the latter contribution always dominates and ultimately yields the Gaussian distribution (\ref{eq:Example2.2}), while the shape of $w(R,\phi)$ is governed by the intrinsic stiffness of the chain for short walks.

In this regime, large deviations from the ideal intersegment angle $\beta$ are suppressed by the angular bias function $p(\theta)$, so walks that significantly contribute to $w(R,\phi)$ show small variations of their bond angles $\theta_k$ around $\beta$.  While these variations may significantly alter the total turning angle $\phi$ of the persistent walk, they usually have little effect on the end-to-end distance $R$, at least when the preferred configuration is mostly stretched (i.~e., $N\beta$ is small).  The situation then resembles the elastic response of a thin rod (or, in the case of chiral bias $\beta \neq0$, a circular spring), with one end fixed at the origin:  While easily yielding to bending forces, the rod (spring) is resilient to compression, and resists changes of its length.  Under applied force, the condition of constant curvature thus requires that the rod bends into a circular arc of fixed length $\Lambda$, and a simple geometrical argument shows that the free end of the rod moves on the curve $R = \Lambda\sin\phi/\phi$.  The shape of this curve is displayed in Figure~\ref{fig:Spring}.

A numerical calculation of the probability distribution function $w(R,\phi)$ for the aforementioned angular bias $p(\theta)$ confirms these deliberations.  In Figure~\ref{fig:Chiral}, we present $w(R,\phi)$ for walks of length $N=4$, $N=7$, and $N=10$.  For the short walk, $w(R,\phi)$ is concentrated along a narrow groove whose shape is reminiscent of the geometrical curve presented in Figure~\ref{fig:Spring}, while for the extended chain, the distribution is largely of Gaussian shape, albeit shifted from the origin, as predicted in (\ref{eq:Example2.5}).  In the transitory case $N=7$, the end-to-end distribution function conspicuously combines the typical features of both limiting cases, as both the central hump and a protruding ridge are clearly visible.  (Note that the isolated spike in the center of the left image is not a numerical artifact; the actual distribution for $N=4$, like the corresponding uncorrelated Pearson walk (Figure~\ref{fig:pdf}) contains a logarithmic singularity at the origin.  In the calculation on display, we chose to truncate the Fourier--Bessel series (\ref{eq:Theory4.8}) after $k = 10^4$ terms.)

\begin{figure*}
\centerline{\includegraphics[draft=false,width=\textwidth]{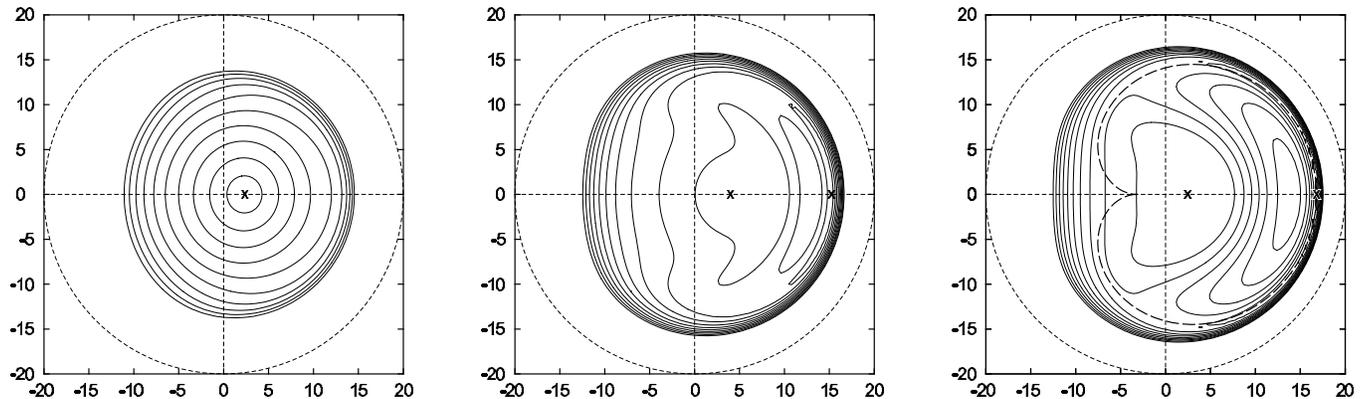}}
\caption{Lines of constant force $F(R,\phi)$ for persistent walks composed of $N=20$ segments.  The angular correlation is of the form $p(\theta) \propto \cos^{2\xi}(\theta/2)$ (\ref{eq:Example2.3}).  With increasing forward bias (left to right: $\xi = 2, 4, 6$), a transition from largely diffusive behavior (a pattern of equidistant concentric circles) to the elastic response of a stiff rod (that is easily bent but resists compression) is observed.  The dashed line in the right plot has been adapted from Figure~\ref{fig:Spring}.
(Spacing of force lines: $0.03\,k_BT/l$.)
\label{fig:Force20}}
\end{figure*}

In a second simulation, we investigate the probability distribution function $w(R,\phi)$ in a persistent walk of $N=20$ segments as a function of the stiffness of the joints, i.~e., the parameter $\xi$ in the angular bias function $p(\theta)$ (\ref{eq:Example2.3}).  In this example, we chose a symmetric distribution ($\beta = 0$) and calculated the average entropic force $\langle F(\mathbf R) \rangle = k_BT |\bm\nabla \log w(\mathbf R)|$ (see Section~\ref{sec:Example1}) for the three stiffness parameters $\xi=2$, $\xi=4$, and $\xi=6$.  (Under these circumstances, convergence of the Fourier--Bessel series (\ref{eq:Theory4.8}) and (\ref{eq:Theory4.11}) was achieved after summing $k=250$ terms.)  Lines of constant force are plotted in Figure~\ref{fig:Force20}.  For the softest chain ($\xi=2$), the pattern resembles a set of equidistant, concentric rings that has been shifted from the origin.  This observation implies that the corresponding distribution $w(R,\phi)$ has approached the long-chain limit (\ref{eq:Example2.2}) characterized by an isotropic harmonic force field $F(\mathbf R)$ directed towards a minimum $\mathbf R_{\min}$ of the free energy $U(\mathbf R) = -k_B T \log w(\mathbf R)$.  Its position is given by the persistence length vector $\mathbf R_{\min} = \langle\mathbf L\rangle_\infty = (l\xi, 0)^T$ (\ref{eq:Example2.5}), and the elastic constants $\epsilon_{jk} = \partial_j\partial_k U(\mathbf R_{\min})$ there read $\epsilon_{xx} = \epsilon_{yy} = 2k_BT/ND_\infty = 2k_BT/[(2\xi+1) Nl^2]$, and $\epsilon_{xy} = 0$.  The numerically determined values $x_{\min} = 2.3224\,l$, $\epsilon_{xx} = 1.522\cdot10^{-2}\,k_BT/l^2$, and $\epsilon_{yy} = 1.438\cdot10^{-2}\,k_BT/l^2$ still deviate considerably from their counterparts in the diffusive limit, $x_{\min} = 2l$ and $\epsilon_{xx} = \epsilon_{yy} = 0.02\,k_BT/l^2$, however.

The force distributions obtained for the stiffer walks with $\xi=4$ and $\xi=6$ bear little resemblance with the diffusive circular pattern observed for $\xi=2$.  For $\xi=4$ (center panel in Figure~\ref{fig:Force20}), the force field is distinctively flat over a wide range of end-to-end vectors $\mathbf R$, and the free energy $U(\mathbf R)$ actually features two local minima $x_< = 3.9808\,l$, $x_> = 15.160\,l$ on the $x$--axis.  The latter represents the maximum of the probability density with $w(x_>,0) \approx 1.91\cdot10^{-3}\,l^{-2}$, but their difference in free energy is only $\Delta U = 0.2336\,k_BT$.  The minima are separated by a shallow maximum near $x_{\max} = 9.15\,l$ ($\Delta U = 0.2846\,k_BT$).  It is instructive to compare the elastic constants $\epsilon_{jk}$ at both minima.  For $x_<$, we find an almost isotropic response ($\epsilon_{xx} = 7.421\cdot10^{-3}\,k_BT/l^2$, $\epsilon_{yy} = 7.009\cdot10^{-3}\,k_BT/l^2$), while the free energy surface near $x_>$ is strongly prolate ($\epsilon_{xx} = 9.529\cdot10^{-2}\,k_BT/l^2$, $\epsilon_{yy} = 2.557\cdot10^{-3}\,k_BT/l^2$).  Clearly, this anisotropy indicates the onset of elastic behavior.

For $\xi=6$ (right panel), elastic response becomes the dominant feature of the distribution which is predominantly concentrated along a narrow arc, while a rather flat plateau in its center still persists.  The maximum of $w(R,\phi)$ has shifted further out to $x_> = 16.790\,l$.  (An extremely shallow local minimum at $x_< = 2.4924\,l$ ($\Delta U = 1.222\,k_BT$) with a barrier of only $0.009\,k_BT$ is contained within the plateau structure.)  The anisotropy of the free energy surface near $x_>$ has become even more marked, where the bending mode ($\epsilon_{yy} = 3.232\cdot10^{-3}\,k_BT/l^2$) is almost 75 times softer than the response towards compression ($\epsilon_{xx} = 2.405\cdot10^{-1}\,k_BT/l^2$).  Empirically, the shape of the trough structure in $U(\mathbf R)$ is accurately reproduced by the elastic spring model (with $\Lambda =Nl$) put forward in Figure~\ref{fig:Spring}, if we allow for a displacement of its origin so that the curve $R(\phi) = Nl\sin\phi/\phi$ passes through $x_>$ (dashed line in Figure~\ref{fig:Force20}).  The elastic compressive response is limited to small deviations from the equilibrium point $x_>$:  For applied loads greater than a critical force $F_{\mathrm{crit}} = 0.195\,k_BT/l$ (at $x_{\mathrm{crit}} = 13.90\,l$), the stretched chain will buckle and collapse into the more globular shapes making up the plateau.  This again resembles the behavior of an elastic rod under stress \cite{Feynman1964a}.

\section{Discussion}
\label{sec:Diss}

In this article, we examined persistent random walks in two dimensions, where the relative orientation of successive segments is correlated through an angular bias function $p(\theta)$.  We found that the eigenfunction expansion into a Fourier--Bessel series provides a convenient and efficient means to calculate the end-to-end probability distribution function $w(\mathbf R)$ encountered in a planar random walk except for extremely short chains.  Related series expansions also cover all derivatives and moments of $w(\mathbf R)$.  We used Pearson's original random walk problem to assess the numerical aspects of the Fourier--Bessel series and found that the number of sum terms required to achieve a desired accuracy initially sharply decreases with the chain length $N$ and, for very extended walks, grows like $(N\log N)^{1/2}$.  This property renders the method attractive in studying the properties of chains of intermediate length whose distribution function $w(\mathbf R)$ may strongly deviate from the diffusive behavior expected in the limit $N\rightarrow \infty$.  Even for uncorrelated planar walks, the singularities contained in $w_P(R)$ cause conspicuous features in the probability distribution and force curves that only gradually fade away as $N$ increases.  Naturally, correlated walks generate a much richer set of phenomena that we only have begun to explore in this work.  Here, we briefly studied ``chiral'' walks with asymmetric angular bias $p(\theta)$ that have a tendency to loop, and inquired into the dependence of $w(\mathbf R)$ on the stiffness of the intersegment joints.  Clear evidence of a transition from diffusive character to a form of elastic response familiar from the elastomechanics of a stiff rod, was obtained.

The latter observation raises questions that deserve further examination.  It appears worthwhile to study the dependence of the crossover from diffusive to elastic behavior as a function of $N$ and the bias $p(\theta)$.  Characteristic parameters of the distribution, like the location of the minima $\mathbf R_{\min}$ of the free energy, the elastic moduli $\epsilon_{jk}$ and the critical force $F_{\mathrm{crit}}$, may obey scaling laws.  The existence of several minima of $U(\mathbf R)$ suggests that the average extension of a stiff chain under an applied external force shows hysteresis.  The formalism developed in this article directly applies to these problems.

Clearly, for many physical applications, e.~g., in polymer theory \cite{Flory1969a}, realistic models will require the random walk to take place in three spatial dimensions.  While the analysis of the uncorrelated random walk problem encounters no major obstacles \cite{Barakat1973a,Treloar1946a}, the extension of the eigenfunction expansion technique to persistent walks in three-dimensional space presents a challenging problem, since the angular bias $p(\theta,\phi)$ then generally will describe the relative orientation of a segment and its \textsl{two} predecessors, unless $p(\theta,\phi)$ is independent of the dihedral angle $\phi$.  Alternative methods, like the transfer operator formalism \cite{Livadaru2002a}, may prove to be better suited for the numerical determination of the probability density function $w(\mathbf R)$ in this case.

\acknowledgments

The author appreciates useful discussions with H.--J.~Kreuzer.  This work was financially supported by the Alexander von Humboldt foundation and the Killam trust.

\vskip 0.5cm plus1cm
\centerline{*\ \ *\ \ *}
\vskip 0.5cm plus1cm

\appendix

\section{The projection theorem}
\label{sec:Projection}

This appendix provides a brief mathematical derivation of the projection theorem for Bessel functions which forms the basis of the Fourier--Bessel expansions derived in Section~\ref{sec:Theory4}.  If $(R,\phi)$ denotes the location of the chain end in polar coordinates for a walk consisting of $N$ steps of individual lengths $\{l_1,\ldots,l_N\}$ characterized by absolute angles $\{\tilde\theta_1, \ldots, \tilde\theta_N\}$ (see Figure~\ref{fig:Angles}), this proposition states that:
\begin{multline}
\label{eq:Proj1}
{\rm e}^{{\rm i}\mu\phi}\, J_\mu\left(z\, \frac R\Lambda \right) = \sum_{\lambda_1 + \ldots + \lambda_N = \mu} {\rm e}^{{\rm i} (\lambda_1\tilde\theta_1 + \ldots + \lambda_N\tilde\theta_N)} \\ \times\, J_{\lambda_1}\left(z\,l_1/\Lambda \right) \cdot\ldots\cdot J_{\lambda_N} \left(z\,l_N/\Lambda \right) \,,
\end{multline}
where $\Lambda=l_1+\ldots+l_N$ represents the contour length of the walk, $\mu\in{\mathbb Z}$ is an integer, and $z\in {\mathbb C}$ denotes an arbitrary complex number.  (For the case of equal step lengths $l_\nu = l$ primarily considered in this article, the arguments of the Bessel functions simplify to $zR/Nl$ on the l.h.s.\ and $z/N$ throughout the r.h.s.\ of this equation, respectively.)

We now sketch the proof of (\ref{eq:Proj1}).  From Figure~\ref{fig:Angles}, it is clear that the projection $\Lambda_x$ of the end-to-end vector on the $x$--axis is simply the sum of the correponding projections of the steps that make up the walk: $R\cos\phi = \sum_\nu l_\nu\cos\tilde\theta_\nu$ ($\nu=1,\ldots,N$).  It is convenient to perform this projection also in an alternative reference frame which is rotated by the angle $\alpha$, i.~e.\ with respect to an arbitrary direction.  Since this operation does not affect the lengths involved and shifts the angles $\phi$ and $\tilde\theta_\nu$ merely by the constant value $\alpha$, we find the relation $R\cos(\phi-\alpha) = \sum_\nu l_\nu \cos(\tilde\theta_\nu - \alpha)$, or in exponential form:
\begin{equation}
\label{eq:Proj2}
{\rm e}^{{\rm i} u R\cos(\phi-\alpha)} = \prod_{\nu=1}^N {\rm e}^{{\rm i}u l_\nu \cos(\tilde\theta_\nu - \alpha)} \;,
\end{equation}
for all $u\in{\mathbb C}$ and $\alpha\in {\mathbb R}$.  We employ the generating function expansion for the Bessel functions $J_\nu(u)$, ${\rm e}^{{\rm i}u\cos\alpha} = \sum_k {\rm i}^k {\rm e}^{{\rm i}k\alpha } J_k(u)$ (\ref{eq:Theory2.11}) on both sides of (\ref{eq:Proj2}) with $u = z/\Lambda$ to obtain the equality:
\begin{multline}
\label{eq:Proj3}
\sum_\lambda \left({\rm i}\,{\rm e}^{-{\rm i} \alpha}\right)^\lambda {\rm e}^{{\rm i}\lambda \phi} J_\lambda\left( z\,\frac R\Lambda \right) = \! \sum_{\lambda_1,\ldots,\lambda_N} \! \left({\rm i}\,{\rm e}^{-{\rm i} \alpha}\right)^{\lambda_1 + \ldots + \lambda_N} \\ \times \, {\rm e}^{{\rm i}(\lambda_1\tilde\theta_1 + \ldots + \lambda_N\tilde\theta_N)} J_{\lambda_1}(z\,l_1/\Lambda) \cdot\ldots\cdot J_{\lambda_N}(z\,l_N/\Lambda) \;.
\end{multline}
The Fourier decomposition of (\ref{eq:Proj3}) with respect to the angle $\alpha$, i.~e.\ an integration $\int_{0}^{2\pi} {\rm d}\alpha \, {\rm e}^{{\rm i}\mu \alpha} \cdot\ldots$, then immediately yields the projection theorem (\ref{eq:Proj1}).

\section{Evaluation of moments}
\label{sec:Moments}

The Fourier--Bessel series (\ref{eq:Theory4.8}) for $w(R,\phi)$ permits fast and accurate numerical evaluation of all moments of the probability distribution in the persistent planar walk.  For this two-dimensional problem, the moments are averages of the general form $\langle L_x^jL_y^k \rangle_N$. Since it is easily verified that they may be expressed as a finite combination of the complex averages $\langle R^{m+2n} {\rm e}^{{\rm i}m\phi} \rangle_N$, where $j+k = m+2n $ and $m,n  \in {\mathbb N}_0$, we prefer to use the radial definition, as it leads to shorter expressions.  Examples are encountered in Section~\ref{sec:Theory1}, where we calculated the average position $\langle L_x +{\rm i}L_y\rangle_N = \langle R {\rm e}^{{\rm i}\phi} \rangle_N$ (\ref{eq:Theory2.1}), corresponding to $m=1$ and $n=0$, and the mean square displacement $\langle L_x^2 + L_y^2\rangle_N = \langle R^2 \rangle_N$ (\ref{eq:Theory2.7}), where $m=0$ and $n=1$.  (We also note in advance that for $m=n=0$, conservation of probability demands $\langle R^0 \rangle_N = 1$.)

The determination of the radial moments involves an integral over Bessel functions that apparently is not listed in the major tables \cite{Gradshteyn1980a}.  Hence, we present a short sketch of its derivation.  We start out with a recursion formula stated by Watson \cite{Watson1944a}.  Introducing the notation:
\begin{equation}
\label{eq:Mom1}
I_{mk}^{(n)} = \int_0^{z_{mk}} \negmedspace {\rm d}u\, u^{m+2n+1} J_m(u) \;,
\end{equation}
where $m,n \geq 0$ are integers, and $z_{mk}$ denotes the $k$.th zero of the Bessel function $J_m(u)$, this recurrence reads:
\begin{equation}
\label{eq:Mom2}
I_{mk}^{(n)} = z_{mk}^{m+2n+1} J_{m+1}(z_{mk}) - 4n(m+n) I_{mk}^{(n-2)} \;.
\end{equation}
Since the recursion terminates for $n=0$, we may add up the terms and rearrange them as the product of $J_{m+1}(z_{mk})$ with a polynomial in the zero $z_{mk}$:
\begin{equation}
\label{eq:Mom3}
I_{mk}^{(n)} = J_{m+1}(z_{mk}) \sum_{\nu=0}^n \frac{(-4)^{n-\nu}n!(m+n)!}{\nu!(m+\nu)!}\, z_{mk}^{m+2\nu+1} \;.
\end{equation}

From the series expansion (\ref{eq:Theory4.8}), we now calculate the radial moments $\langle R^{m+2n} {\rm e}^{{\rm i}m\phi} \rangle_N$.  Since the angular integration is trivial, using the abbreviation (\ref{eq:Mom1}) in the first step we obtain:
\begin{equation}
\label{eq:Mom4}
\begin{aligned}
\left\langle \cdots \right\rangle_N &= \int_0^{Nl} \negmedspace {\rm d}R\, R^{m+2n+1} \angint {\rm d}\phi \,{\rm e}^{{\rm i}m\phi} w(R,\phi) \\
&= 2 (Nl)^{m+2n} \sum_{k=1}^\infty \frac{\left[{\cal Z}(z_{mk}/N)^N\right]_{0m} I_{mk}^{(n)}}{z_{mk}^{m+2n+2} J_{m+1}(z_{mk})^2} \;,
\end{aligned}
\end{equation}
where the elements of the matrix ${\cal Z}$ are given by ${\cal Z}_{jl} = J_{l-j}(z_{mk}/N)p_l$ (see Section~\ref{sec:Theory4}).  Using (\ref{eq:Mom3}), we rearrange this sum into a series over powers of $z_{mk}$.  It is convenient to define a set of auxiliary sums $B_{mr}^{(N)}$ involving the zeroes of the Bessel functions $J_m(u)$:
\begin{equation}
\label{eq:Mom5}
B_{mr}^{(N)} = \sum_{k=1}^\infty \frac{\left[{\cal Z}(z_{mk}/N)^N\right]_{0m}}{z_{mk}^{2r+1} J_{m+1}(z_{mk})} \;.
\end{equation}
The final result for the radial moments then can be cast in the following form:
\begin{multline}
\label{eq:Mom6}
\left\langle R^{m+2n} {\rm e}^{{\rm i}m\phi} \right\rangle_N = 2^{2n+1} (Nl)^{m+2n} \\
\times\, \sum_{\nu=0}^n \frac{(-1)^{n-\nu}n!(m+n)!}{2^{2\nu}\nu!(m+\nu)!} \,B_{m,n-\nu}^{(N)} \;.
\end{multline}
For not too small $N$, the series $B_{mr}^{(N)}$ (\ref{eq:Mom5}) converge rapidly, thus permitting efficient numerical evaluation of the moments using (\ref{eq:Mom6}).  (For practical purposes, we note that the average $\langle R^{m+2n} {\rm e}^{{\rm i}m\phi} \rangle_N$ (\ref{eq:Mom6}) is a function of the angular Fourier coefficients $p_0,p_{\pm1},\ldots,p_{\pm(m+n)}$ only, as the method of direct evaluation exposed in Section~\ref{sec:Theory1} shows.  In the calculation of $B_{mr}^{(N)}$ (\ref{eq:Mom5}), it is thus convenient to set $p_\mu = 0$ for $|\mu| > m+r$.)

Finally, we present several examples to illustrate our result for the moments of $w(R,\phi)$.  For $n=0$, the series (\ref{eq:Mom6}) reduces to a single term:  $\langle (L_x + {\rm i}L_y)^m \rangle_N = 2 (Nl)^m B_{m0}^{(N)}$.  In the case $m=0$,  conservation of probability then requires $B_{00}^{(N)} = 1/2$, irrespective of $N$ and $p(\theta)$. Similarly, comparison with (\ref{eq:Theory1.1}) yields for $m=1$:
\begin{equation}
\label{eq:Mom7}
B_{10}^{(N)} = \frac{p_1(1-p_1^N)}{2N (1-p_1)} \;.
\end{equation}
We may use these relations to keep track of the accuracy in the summation of the Fourier-Bessel series (\ref{eq:Theory4.8}).  Similarly, from (\ref{eq:Mom6}) we obtain all purely radial moments $\langle R^{2n}\rangle_N$ of even order.  They read for $n\leq 3$:
\begin{align}
\label{eq:Mom8a}
\left\langle R^2\right\rangle_N = & (Nl)^2 \left( 1 - 8 B_{01}^{(N)} \right) \;, \\
\label{eq:Mom8b}
\left\langle R^4\right\rangle_N = & (Nl)^4 \left( 1 - 32 B_{01}^{(N)} + 128 B_{02}^{(N)} \right) \;, \\
\label{eq:Mom8c}
\left\langle R^6\right\rangle_N = & (Nl)^6 \left( 1 - 72 B_{01}^{(N)} + 1152 B_{02}^{(N)} - 4608 B_{03}^{(N)}  \right) \;,
\end{align}
where we already inserted $B_{00}^{(N)} = 1/2$.  Clearly, an expression for $B_{01}^{(N)}$ in terms of the angular bias Fourier components $p_{\pm1}$ akin to (\ref{eq:Mom7}) is available from our earlier result for the mean square displacement (\ref{eq:Theory1.7}).  From there, we also infer that $\langle R^2 \rangle_N$ in the limit $N\rightarrow\infty$ only grows linearly with $N$ (\ref{eq:Theory2.8}).  Hence, asymptotically the Bessel sum coefficient $B_{01}^{N}$ in (\ref{eq:Mom8a}) approaches the value $B_{01}^{\infty} = \frac18$.  Similar considerations for the higher moments in conjunction with (\ref{eq:Mom5}) then show that:
\begin{equation}
\label{eq:Mom9}
B_{0r}^{(\infty)} = \sum_{k=1}^\infty \frac1{z_{0k}^{2r+1} J_1(z_{0k})} = \frac{a_r}{2^{2r+1}(r!)^2} \;,
\vphantom{\int_{\int_\int}}
\end{equation}
where the coefficients $a_r$ form a sequence of integers: $a_0=1$, $a_1=1$, $a_2=3$, $a_3=19$, etc.

\bibliography{Persist}

\begin{thebibliography}{25}
\expandafter\ifx\csname natexlab\endcsname\relax\def\natexlab#1{#1}\fi
\expandafter\ifx\csname bibnamefont\endcsname\relax
  \def\bibnamefont#1{#1}\fi
\expandafter\ifx\csname bibfnamefont\endcsname\relax
  \def\bibfnamefont#1{#1}\fi
\expandafter\ifx\csname citenamefont\endcsname\relax
  \def\citenamefont#1{#1}\fi
\expandafter\ifx\csname url\endcsname\relax
  \def\url#1{\texttt{#1}}\fi
\expandafter\ifx\csname urlprefix\endcsname\relax\def\urlprefix{URL }\fi
\providecommand{\bibinfo}[2]{#2}
\providecommand{\eprint}[2][]{\url{#2}}

\bibitem[{\citenamefont{Shlesinger and West}(1984)}]{Shlesinger1984a}
\bibinfo{editor}{\bibfnamefont{M.F.}~\bibnamefont{Shlesinger}} \bibnamefont{and}
  \bibinfo{editor}{\bibfnamefont{B.J.}~\bibnamefont{West}}, eds.,
  \emph{\bibinfo{title}{Random Walks and Their Applications in the Physical and
  Biological Sciences}}, no. \bibinfo{number}{109} in \bibinfo{series}{AIP
  Conference Proceedings} (\bibinfo{publisher}{American Institute of Physics},
  \bibinfo{address}{New York}, \bibinfo{year}{1984}).

\bibitem[{\citenamefont{Weiss}(1994)}]{Weiss1994a}
\bibinfo{author}{\bibfnamefont{G.H.}~\bibnamefont{Weiss}},
  \emph{\bibinfo{title}{Aspects and Applications of the Random Walk}}
  (\bibinfo{publisher}{North--Holland}, \bibinfo{address}{Amsterdam},
  \bibinfo{year}{1994}).

\bibitem[{\citenamefont{Pearson}(1905)}]{Pearson1905a}
\bibinfo{author}{\bibfnamefont{K.}~\bibnamefont{Pearson}},
  \bibinfo{journal}{Nature} \textbf{\bibinfo{volume}{72}}, \bibinfo{pages}{294}
  (\bibinfo{year}{1905}).

\bibitem[{\citenamefont{Masoliver et~al.}(1993)\citenamefont{Masoliver,
  Porr{\`a}, and Weiss}}]{Masoliver1993a}
\bibinfo{author}{\bibfnamefont{J.}~\bibnamefont{Masoliver}},
  \bibinfo{author}{\bibfnamefont{J.M.}~\bibnamefont{Porr{\`a}}},
  \bibnamefont{and} \bibinfo{author}{\bibfnamefont{G.H.}~\bibnamefont{Weiss}},
  \bibinfo{journal}{Physica~A} \textbf{\bibinfo{volume}{193}},
  \bibinfo{pages}{469} (\bibinfo{year}{1993}).

\bibitem[{\citenamefont{Tojo and Argyrakis}(1996)}]{Tojo1996a}
\bibinfo{author}{\bibfnamefont{C.}~\bibnamefont{Tojo}} \bibnamefont{and}
  \bibinfo{author}{\bibfnamefont{P.}~\bibnamefont{Argyrakis}},
  \bibinfo{journal}{Phys.~Rev.~E} \textbf{\bibinfo{volume}{54}},
  \bibinfo{pages}{58} (\bibinfo{year}{1996}).

\bibitem[{\citenamefont{Larralde}(1997)}]{Larralde1997a}
\bibinfo{author}{\bibfnamefont{H.}~\bibnamefont{Larralde}},
  \bibinfo{journal}{Phys.~Rev.~E} \textbf{\bibinfo{volume}{56}},
  \bibinfo{pages}{5004} (\bibinfo{year}{1997}).

\bibitem[{\citenamefont{Wu et~al.}(2000)\citenamefont{Wu, Li, Springer, and
  Neill}}]{Wu2000a}
\bibinfo{author}{\bibfnamefont{H.}~\bibnamefont{Wu}},
  \bibinfo{author}{\bibfnamefont{B.L.}~\bibnamefont{Li}},
  \bibinfo{author}{\bibfnamefont{T.A.}~\bibnamefont{Springer}}, \bibnamefont{and}
  \bibinfo{author}{\bibfnamefont{W.H.}~\bibnamefont{Neill}},
  \bibinfo{journal}{Ecol. Modell.} \textbf{\bibinfo{volume}{132}},
  \bibinfo{pages}{115} (\bibinfo{year}{2000}).

\bibitem[{\citenamefont{Barakat}(1974)}]{Barakat1974a}
\bibinfo{author}{\bibfnamefont{R.}~\bibnamefont{Barakat}},
  \bibinfo{journal}{Optica Acta} \textbf{\bibinfo{volume}{21}},
  \bibinfo{pages}{921} (\bibinfo{year}{1974}).

\bibitem[{\citenamefont{Shmueli and Weiss}(1990)}]{Shmueli1990a}
\bibinfo{author}{\bibfnamefont{U.}~\bibnamefont{Shmueli}} \bibnamefont{and}
  \bibinfo{author}{\bibfnamefont{G.H.}~\bibnamefont{Weiss}},
  \bibinfo{journal}{J.~Am.~Stat.~Soc.} \textbf{\bibinfo{volume}{85}},
  \bibinfo{pages}{6} (\bibinfo{year}{1990}).

\bibitem[{\citenamefont{Shmueli and Weiss}(1995)}]{Shmueli1995a}
\bibinfo{author}{\bibfnamefont{U.}~\bibnamefont{Shmueli}} \bibnamefont{and}
  \bibinfo{author}{\bibfnamefont{G.H.}~\bibnamefont{Weiss}},
  \emph{\bibinfo{title}{Introduction to Crystallographic Statistics}}
  (\bibinfo{publisher}{Oxford Science Publications}, \bibinfo{address}{Oxford},
  \bibinfo{year}{1995}).

\bibitem[{\citenamefont{Flory}(1969)}]{Flory1969a}
\bibinfo{author}{\bibfnamefont{P.J.}~\bibnamefont{Flory}},
  \emph{\bibinfo{title}{Statistical Mechanics of Chain Molecules}}
  (\bibinfo{publisher}{Interscience}, \bibinfo{address}{New York},
  \bibinfo{year}{1969}).

\bibitem[{\citenamefont{Nossal and Weiss}(1974{\natexlab{a}})}]{Nossal1974a}
\bibinfo{author}{\bibfnamefont{R.}~\bibnamefont{Nossal}} \bibnamefont{and}
  \bibinfo{author}{\bibfnamefont{G.H.}~\bibnamefont{Weiss}},
  \bibinfo{journal}{J.~Stat.~Phys.} \textbf{\bibinfo{volume}{10}},
  \bibinfo{pages}{245} (\bibinfo{year}{1974}{\natexlab{a}}).

\bibitem[{\citenamefont{Nossal and Weiss}(1974{\natexlab{b}})}]{Nossal1974b}
\bibinfo{author}{\bibfnamefont{R.}~\bibnamefont{Nossal}} \bibnamefont{and}
  \bibinfo{author}{\bibfnamefont{G.H.}~\bibnamefont{Weiss}},
  \bibinfo{journal}{J.~Theor.~Biol.} \textbf{\bibinfo{volume}{47}},
  \bibinfo{pages}{103} (\bibinfo{year}{1974}{\natexlab{b}}).

\bibitem[{\citenamefont{Nossal}(1983)}]{Nossal1983a}
\bibinfo{author}{\bibfnamefont{R.}~\bibnamefont{Nossal}},
  \bibinfo{journal}{J.~Stat.~Phys.} \textbf{\bibinfo{volume}{30}},
  \bibinfo{pages}{391} (\bibinfo{year}{1983}).

\bibitem[{\citenamefont{Barakat}(1973)}]{Barakat1973a}
\bibinfo{author}{\bibfnamefont{R.}~\bibnamefont{Barakat}},
  \bibinfo{journal}{J.~Phys.~A} \textbf{\bibinfo{volume}{6}},
  \bibinfo{pages}{796} (\bibinfo{year}{1973}).

\bibitem[{\citenamefont{Weiss and Shmueli}(1987)}]{Weiss1987a}
\bibinfo{author}{\bibfnamefont{G.H.}~\bibnamefont{Weiss}} \bibnamefont{and}
  \bibinfo{author}{\bibfnamefont{U.}~\bibnamefont{Shmueli}},
  \bibinfo{journal}{Physica~A} \textbf{\bibinfo{volume}{146}},
  \bibinfo{pages}{641} (\bibinfo{year}{1987}).

\bibitem[{\citenamefont{Kluyver}(1906)}]{Kluyver1906a}
\bibinfo{author}{\bibfnamefont{J.C.}~\bibnamefont{Kluyver}},
  \bibinfo{journal}{Proc. Section of Sci.~K.~Acad.\ van Wet.~te Amsterdam}
  \textbf{\bibinfo{volume}{8}}, \bibinfo{pages}{341} (\bibinfo{year}{1906}).

\bibitem[{\citenamefont{Gradshteyn and Ryzhik}(1980)}]{Gradshteyn1980a}
\bibinfo{author}{\bibfnamefont{I.S.}~\bibnamefont{Gradshteyn}} \bibnamefont{and}
  \bibinfo{author}{\bibfnamefont{I.M.}~\bibnamefont{Ryzhik}},
  \emph{\bibinfo{title}{Table of Integrals, Series, and Products}}
  (\bibinfo{publisher}{Academic Press}, \bibinfo{address}{San Diego},
  \bibinfo{year}{1980}).

\bibitem[{\citenamefont{Kramers and Wannier}(1941)}]{Wannier1941a}
\bibinfo{author}{\bibfnamefont{H.A.}~\bibnamefont{Kramers}} \bibnamefont{and}
  \bibinfo{author}{\bibfnamefont{G.H.}~\bibnamefont{Wannier}},
  \bibinfo{journal}{Phys.~Rev.} \textbf{\bibinfo{volume}{60}},
  \bibinfo{pages}{252} (\bibinfo{year}{1941}).

\bibitem[{\citenamefont{Montroll}(1947)}]{Montroll1947a}
\bibinfo{author}{\bibfnamefont{E.W.}~\bibnamefont{Montroll}},
  \bibinfo{journal}{Ann.~Math.~Stat.} \textbf{\bibinfo{volume}{18}},
  \bibinfo{pages}{18} (\bibinfo{year}{1947}).

\bibitem[{\citenamefont{Watson}(1944)}]{Watson1944a}
\bibinfo{author}{\bibfnamefont{G.N.}~\bibnamefont{Watson}},
  \emph{\bibinfo{title}{A Treatise on the Theory of {B}essel Functions}}
  (\bibinfo{publisher}{Cambridge University Press},
  \bibinfo{address}{Cambridge}, \bibinfo{year}{1944}).

\bibitem[{\citenamefont{Sommerfeld}(1949)}]{Sommerfeld1949a}
\bibinfo{author}{\bibfnamefont{A.}~\bibnamefont{Sommerfeld}},
  \emph{\bibinfo{title}{Partial Differential Equations in Physics}}
  (\bibinfo{publisher}{Academic Press}, \bibinfo{address}{New York},
  \bibinfo{year}{1949}).

\bibitem[{\citenamefont{Treloar}(1946)}]{Treloar1946a}
\bibinfo{author}{\bibfnamefont{L.R.G.}~\bibnamefont{Treloar}},
  \bibinfo{journal}{Trans.~Faraday~Soc.} \textbf{\bibinfo{volume}{42}},
  \bibinfo{pages}{77} (\bibinfo{year}{1946}).

\bibitem[{\citenamefont{Feynman et~al.}(1964)\citenamefont{Feynman, Leighton,
  and Sands}}]{Feynman1964a}
\bibinfo{author}{\bibfnamefont{R.P.}~\bibnamefont{Feynman}},
  \bibinfo{author}{\bibfnamefont{R.B.}~\bibnamefont{Leighton}}, \bibnamefont{and}
  \bibinfo{author}{\bibfnamefont{M.}~\bibnamefont{Sands}},
  \emph{\bibinfo{title}{The Feynman Lectures on Physics}}
  (\bibinfo{publisher}{Addison--Wesley}, \bibinfo{address}{Reading, Mass.},
  \bibinfo{year}{1964}), vol.~\bibinfo{volume}{2}, chap.~\bibinfo{chapter}{38}.

\bibitem[{\citenamefont{Livadaru et~al.}(2002)\citenamefont{Livadaru, Netz, and
  Kreuzer}}]{Livadaru2002a}
\bibinfo{author}{\bibfnamefont{L.}~\bibnamefont{Livadaru}},
  \bibinfo{author}{\bibfnamefont{R.R.}~\bibnamefont{Netz}}, \bibnamefont{and}
  \bibinfo{author}{\bibfnamefont{H.J.}~\bibnamefont{Kreuzer}},
  \bibinfo{journal}{Macromolecules}  (\bibinfo{year}{2002}),
  \bibinfo{note}{(submitted)}.

\end{thebibliography}
\end{document}